\documentclass[12pt,a4paper]{article}
\usepackage{jheppub}
\usepackage[english]{babel}
\usepackage{graphicx}
\usepackage[titletoc]{appendix}
\usepackage{xspace}
\usepackage{hyperref}
\usepackage[utf8]{inputenc}
\usepackage{listings}
\usepackage{xcolor}
\usepackage{longtable}
\usepackage{pdflscape}
\usepackage{float}

\definecolor{codegreen}{rgb}{0,0.6,0}
\definecolor{codegray}{rgb}{0.5,0.5,0.5}
\definecolor{codepurple}{rgb}{0.58,0,0.82}
\definecolor{backcolour}{rgb}{0.95,0.95,0.92}

\lstdefinestyle{mystyle}{
    backgroundcolor=\color{backcolour},   
    commentstyle=\color{codegreen},
    keywordstyle=\color{magenta},
    numberstyle=\tiny\color{codegray},
    stringstyle=\color{codepurple},
    basicstyle=\ttfamily\footnotesize,
    breakatwhitespace=false,         
    breaklines=true,                 
    captionpos=b,                    
    keepspaces=true,                 
    numbers=left,                    
    numbersep=5pt,                  
    showspaces=false,                
    showstringspaces=false,
    showtabs=false,                  
    tabsize=2
}
  
\title{HL-LHC Analysis With ROOT \\[1em]
 \large ROOT Project Input to the HL-LHC Computing Review Stage 2}
\collaboration{The ROOT Team, September 2021}
\emailAdd{rootdev@cern.ch}

\author[a]{Bertrand Bellenot,}
\author[a]{Jakob Blomer,}
\author[b]{Philippe Canal,}
\author[a]{Olivier Couet,}
\author[a]{Javier Lopez Gomez,}
\author[a]{Bernhard Manfred Gruber,}
\author[a]{Enrico Guiraud,}
\author[a]{Jonas Hahnfeld,}
\author[d]{Sergey Linev,}
\author[a]{Lorenzo Moneta,}
\author[a]{Axel Naumann,}
\author[a]{Vincenzo Eduardo Padulano,}
\author[a]{Jonas Rembser,}
\author[c]{Alja Mrak Tadel,}
\author[c]{Matevz Tadel,}
\author[a]{Enric Tejedor and }
\author[e]{Vassil Vassilev}

\affiliation[a]{CERN, Geneva, Switzerland}
\affiliation[b]{Fermi National Accelerator Laboratory, Batavia, USA}
\affiliation[c]{University of California San Diego, USA}
\affiliation[d]{GSI, Darmstadt, Germany}
\affiliation[e]{Princeton University, USA}

\date{September 2021}

\makeatletter
\gdef\@fpheader{}
\makeatother

\begin{document}
\maketitle

\tableofcontents

\section{Introduction}

The HL-LHC analysis is expected \cite{HLLHCAnalysisWorkshop} to be similar to Run 3 analyses, with more events (data and simulation),
possibly higher number of samples (multiple background channels,
side-band channels from data, comparing more collision event generators,
higher number of data skims), higher complexity of fits due to more
precise models with higher parameter counts, and more common usage of
Machine Learning to improve the physics performance of analysis
algorithms.
As a consequence of more complex fits and the usage of
machine learning, ROOT expects a further increase in the usage of GPUs
and possibly other accelerators such as TPUs.

ROOT is preparing for HL-LHC analyses through a series of R\&D tasks
started several years ago, to be able to address these challenges in the
form of an integrated solution.
It will do this in time for HL-LHC,
given the resources and ROOT's experience with a sustainable feedback
and development velocity, and the adoption rate in the physics
community.
Several of these undertakings are massive (in ROOT's scale)
with respect to the required effort and expected impact.

Until now, ROOT has provided the core ingredients for virtually all LHC
analyses.
Because of its role as "integrated analysis foundation
library", much of ROOT can be used by ROOT itself, increasing the impact
of its own developments:
all of ROOT's features are optimized with
respect to I/O; many make use of ROOT's visualization abilities;
fundamental features such as just-in-time compilation and network calls
are enabling fundamentally new features such as remote event displays.

Being a central element of the HEP software ecosystem, ROOT sees many
contributions, from volunteer, occasional improvements to
institutionalized cooperation.
Whatever the contribution, it will become
part of ROOT only if the contribution is relevant to the community as a
whole.
Or, conversely, any successful contribution to ROOT has a large
impact on the community, multiplied through the tens of thousand
physicists and virtually all HEP experiments, world-wide.

Given the experience and expertise of the ROOT project, ROOT believes
the main HL-LHC analysis challenges to be:

\begin{itemize}

\item
  Data: high-throughput, local and remote data access to reduce the
  impact of analyses' traditional I/O bottleneck.
\item
  Efficiency and simplicity: making it obvious for physicists how to
  write an analysis with faster time-to-results.
\item
  Robustness and correctness: providing results that can be trusted and
  published
\item
  Interoperability: enabling physicists to use the language of choice,
  with the packages of choice, without sacrificing performance or
  simplicity.
\item
  Sustainability and Innovation: investing in solutions that the
  community can benefit from for the next decades.
\end{itemize}

This part of the document introduces the main analysis challenges as
perceived by the ROOT project, alongside their potential remedies with
associated risks.

\section{Analysis Ecosystem}

ROOT continues to be a catalyzer for an analysis ecosystem around it. At
the same time, Python is expected to become the dominating HL-LHC
analysis language, and ROOT is further integrating into a wider Python
ecosystem. ROOT works towards strengthening \emph{both} aspects - by
providing the integrated, fundamental ingredients targeted at HEP that
enable ecosystems to grow around it, and by providing smooth
interoperability to the Python analysis ecosystem in a reusable way for
easy integration with emerging ecosystems such as Julia.

\subsection{Machine Learning}
\label{ch:AE-ML}

Most primary machine learning (ML) interfaces are written in Python,
encouraging the transfer of large amounts of data for instance to GPUs
which then internally run non-Python code. ROOT works towards the
integration with such frameworks, specifically PyTorch and TensorFlow's
generator interface, to have simple and highly efficient user-facing
interaction between ROOT's analysis interface RDataFrame and ML tools.
This benefits from ROOT's ability to optimize the HEP I/O layer.
Examples include ROOT's ability to arrange for in-memory data layout and
alignment that corresponds to on-disk layouts, reducing or removing the
need for CPU cycles when transferring data into GPUs.

Investment in this area (1.5\,FTE over 3 years) is needed for the
analysis community to benefit from ROOT's I/O layer, and to create a
performance and usability benefit so significant that analysts do not
migrate to less performant options due to a perceived increase of
productivity. It is trivial for physicists to find examples for use of
machine learning for instance from data in text format. Without
additional benefits and clear, communicated advantages, large parts of
the community will adopt less efficient (in terms of CPU and
storage-space requirements) approaches simply because of their much
higher online presence. This is an inherent risk of the wider adoption
of Python as the preferred analysis language, with a transformation away
from the single, community-agreed, consistently designed, integrated and
performance-conscious analysis framework, to an ecosystem of literally
thousands of potential building blocks
{[}\href{https://pypi.org/}{\emph{https://pypi.org/}}{]}. Investment in
this area (0.5\,FTE over 4 years) can help reduce the residual risk that
comes with Python becoming the primary analysis language, which is
nowhere as visible as in the context of machine learning, and which we
expect to continue during HL-LHC.

ROOT's role with ML must thus be to integrate external machine learning
libraries with ROOT's I/O and analysis facilities. This requires
expertise in Python and C++, ML, GPUs, as well as code and memory layout
optimization, as an ongoing effort (0.5\,FTE/y). ROOT's R\&D on \href{https://indico.cern.ch/event/999647/}{RNTuple}
(1\,FTE/y over 3 years),
\href{https://github.com/alpaka-group/llama}{LLAMA} (0.5\,FTE/y over 4 years),
\href{https://resources.nvidia.com/events/GTC2020s21588}{cling-CUDA integration} (0.1\,FTE/y over 3 years)
and \href{https://github.com/vgvassilev/clad}{automatic differentiation} (0.5\,FTE/y over 4 years)
are significant building blocks in that respect. First results for
RDataFrame-based training on GPUs are expected in 2022; ML-optimized
RNTuple-to-GPU transfer leveraging above mentioned R\&D will be
implemented by 2024 (0.75\,FTE over 2 years).

The optimization of machine-learning models is an interactive,
user-facing activity. ROOT used to provide a GUI for the visual
inspection of the training quality (for instance versus epoch). This has
many users, but needs to be improved and ported to ROOT's new web-based
GUI, with all its advantages. Work for this is underway (0.25\,FTE/y),
and should complete by 2024.

Dissemination of optimal analysis approaches again underlines the
necessity for ROOT's experts to be trusted and visible in the HEP ML and
analysis community. It requires the continuous investment of a
significant training effort, by said ROOT ML experts (0.25FTE/y).

\subsection{Python}

Python is seeing a constantly increasing adoption rate as language of
choice for analyses. It is important to understand that this is the
\emph{surface} layer: just as bash scripts steer programs for the
computational work, Python code is ideally delegating resource-intense
work to for instance C/C++ libraries. With PyROOT, ROOT provides the
"glue" used by HEP to make HEP's C++ libraries accessible transparently
through Python. With significantly more than ten years of experience,
ROOT has learned how to design C++ interfaces in a way that makes them
good Python interfaces, too. A good example is ROOT's analysis interface
RDataFrame, where virtually identical user code is written, be it in C++
or Python \cite{RDF}. While much of the
(pseudo-) object oriented interfaces match nicely across languages, some
idioms are language-specific and need to be adapted. PyROOT enabled this
through a mechanism called "pythonizations": features added on top of
the default bindings to make ROOT easier to use from Python, or in a
more intuitive way for Python programmers.

PyROOT relies on a project called cppyy \cite{cppyy},
a tool used for instance in computational biology. Cppyy in turn relies
on ROOT's type description system and cling, ROOT's interpreter / JIT,
to dynamically create the binding between Python and C++. This binding
layer needs to evolve with C++, supporting new language features and
data types. Significant expertise is required for this, in the areas of
Python, and in cling and ROOT's type description system, which provide
PyROOT with an answer to questions such as "which methods does
edm::Collection\textless edm::Jet\textgreater{} have". Due to a lack of
past investment, this layer has accumulated significant technical debt.
The current state is hindering gradual evolution with minimal
investment, and \href{https://bitbucket.org/wlav/cppyy/issues/5/requirements-for-use-of-roots-cling}{limits PyROOT's performance and functionality}.
Addressing this would require an investment of 1\,FTE over 4 years.

The experiments' frameworks are aggressively embracing new C++
standards, cling and PyROOT are expected to support new features as
quickly as possible. The continuous investment of 0.5\,FTE/y is necessary
to prevent a reduction of functionality from Python, which in turn can
mean usage migrating away from performant C++ libraries.

ROOT's interoperability with Python libraries such as NumPy is crucial
for high performance data processing. ROOT plans to invest in a C++
container that allows data transfer directly into NumPy arrays. This
work (0.25\,FTE over 1 year) is expected to conclude by 2023.

ROOT remains a C++ program and library; yet - seeing the widespread use
of Python in analyses - needs to ensure that ROOT can be used
"naturally" from Python, in terms of interoperability (data formats such
as NumPy), interface style, Python syntax, object ownership. This
requires investment in existing and future C++ interfaces. It even
stipulates the re-design of interfaces such as RooFit. Without such an
investment, a fair part of the community is expected to migrate to
alternative solutions. The number of available alternatives - even today
- will cause a fragmentation of the community. This reduces the benefits
of central investment; all known solutions have feature limitations,
risking a reduced physics reach for analyses; solutions will not be
easily applicable to other problems due to the expected feature
segregation. To counter these risks, ROOT has invested significantly in
PyROOT and for instance RooFit, to improve ease of use, stability,
performance, support, documentation - in short: to reduce the
physicists' need to migrate away. This effort needs to be maintained (1
FTE/y), for ROOT to be able to continue to compete, and to reduce the
residual risk of community fragmentation. At the same time, competition
needs to be fostered, especially in the Python ecosystem, which sees
much innovation (especially in interfaces) and can be a fruitful,
effective ground for software experiments. This competition is input to
ROOT's evolution, and allows to benchmark ROOT against alternatives.

For several of years now, ROOT has been \href{https://root.cern/install/\#conda}{released also through Conda}.
This gives many Python users trivial access to ROOT. The notion of "ROOT
is C++ and thus awkward to use in Python" needs to be overcome, by
continuing to invest in simple distribution mechanisms and perfect
embedding in Python ecosystems, making ROOT accessible in Python
ecosystems as easily as any other Python packaged C++ library
(continuous 0.1\,FTE/y).

What has been done for Python with PyROOT can be (and has been) done for
other languages, in a similar way. ROOT's C++ interpreter / JIT compiler
cling allows ROOT to create such language bindings dynamically for
languages that might become relevant to HEP analyses in the future.

\subsection{Data Format}

Up until Run3, the LHC experiments used ROOT files and TTrees; several,
mostly small-scale, studies have questioned that and proposed
alternatives. A more complete and in-depth review showed the advantages
of ROOT's file format and TTree \cite{HEPDataFormatComparison}. Nonetheless, these studies and for instance ROOT's
experience with supporting I/O in multi-threaded environments have shown
limitations to TTree. As ROOT's I/O subsystem guarantees backward
compatibility (old ROOT files can be read with new ROOT versions) as
well as forward compatibility (new ROOT files can be read with old ROOT
versions), evolution of TTree is severely limited.

\begin{figure}[H]
\includegraphics[clip, trim=0cm 0cm 0cm 6.5cm, width=1\textwidth]{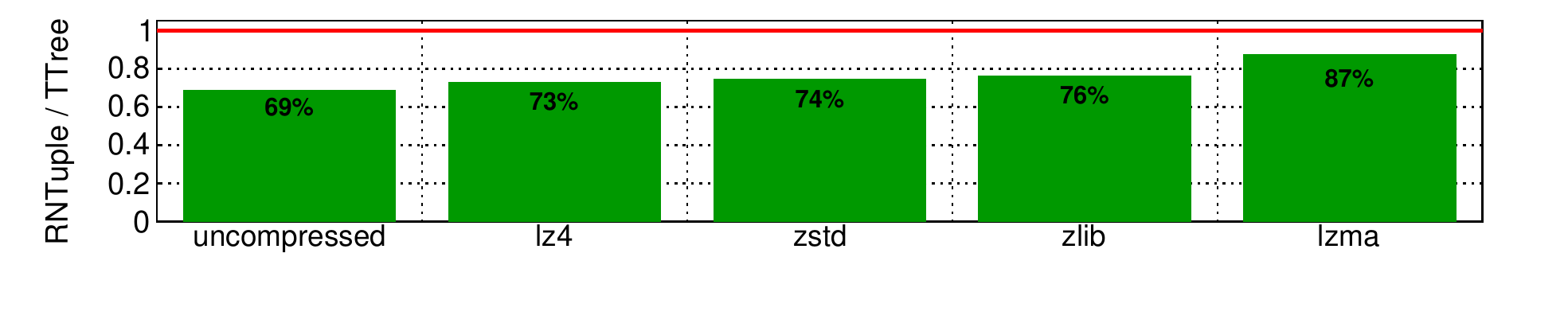}
\caption{Fraction of storage size RNTuple vs. TTree for identical data content (CMS NanoAOD);
lower is better.}
\label{RNTupleStorageEff}
\end{figure}

R\&D on potential benefits of a new data layout for HEP, called RNTuple \cite{RNTuple}, showed improvements to
transfer rate, storage size efficiency (see Fig.~\ref{RNTupleStorageEff}), robustness, and flexibility that
are sufficiently significant to warrant the introduction of a new,
evolved I/O subsystem for HL-LHC, see the \emph{Foundation} part of the ROOT input. ROOT
caters both to frameworks and analysis physicists. It was thus paramount
to make RNTuple work exceptionally well also for analyses. RNTuple is
expected to have a significant performance effect on machine learning
(more than a factor 10 read throughput in training compared to TTree)
and RDataFrame, ROOT's modern analysis interface (factor 2 in read
throughput). These two standard ingredients of analyses are
traditionally I/O limited and will benefit directly from RNTuple.

ROOT expects to provide a tool converting TTrees to RNTuples in 2021
(0.2\,FTE). For several years now, ROOT has been advocating and training
physicists in the use of RDataFrame, also as an abstraction of the
underlying data layout and to ease the migration from TTree to RNTuple.
RDataFrame can process ROOT files containing TTrees just as well as ROOT
files containing RNTuples. ROOT will continue to maintain the TTree
interfaces. Nonetheless we expect to see a migration away from the use
of low level TTree interfaces to RDataFrame, enabling optimizations and
concurrency. After all, the data format will be defined by the
experiments; it is ROOT's task to ensure a smooth migration is possible
for physicists.

Alternative data formats continue to be used in HEP, although it is
unclear whether their adoption is increasing. As analyses are probably
the most agile part of HEP's software environment, they are expected to
continue to try alternatives (HDF5, Parquet, etc), also in function of
the tools and libraries used by analyses \cite{HEPQueryLang}.
ROOT's goal is to preempt such re-formatting, which has consistently
proven as a bottleneck for analyses' agility, preventing smooth
integration of optimizations in the ROOT-part of the analyses, and
increasing the storage needs for analyses. During the coming years, ROOT
will continue to work on providing efficient and easy to use interfaces.
Examples include interfaces to ML \cite{TMVAFastInference}, for transparent use of
multi-core and GPUs \cite{RDF,NewRooFit}, for
distributed computing \cite{DistibRDF}, to
Python \cite{PyROOT}, and for data visualization.

\section{Empowering Physicists}

ROOT's role is to facilitate analyses. Helping physicists get analysis
results quickly, reliably, and with less resources, is what ROOT is
striving to improve, constantly.

\subsection{Analysis as Effective Use of
Luminosity, Efficiency of Physicists, Time to Publication}

There are two drawbacks in high turn-around times of analyses: on one
hand, this simply delays the results, pushing what could be published in
2025 to 2026. This used to be perceived as a mere optimization issue.
Being first or a runner-up seems to be of increasing significance to
funding agencies, the public, and experiments as a whole.

On the other hand, the analysis throughput of physicists is a limited
resource. Making physicists more productive simply means better use of
luminosity. Or in other terms, the analysis efficiency should be
considered as an integral part of the experiments' overall efficiency:
it should not stop at reconstruction.

Increasing the physics reach of a typical PhD analysis should thus be an
important goal for the community. This can be considered as part of the
overall efficiency, and included in optimizations and upgrades. As
detailed below, ROOT has the proven potential to measurably increase the
effectiveness of analyses by factors, at a cost that is shared between
experiments and that is far lower than investment in detector parts.

\subsection{Multi-Platform Support}

Being mostly written in C++, ROOT's code can make efficient use of CPUs.
This brings a certain platform dependence: compiled C++ code will only
work on a given set of architectures; optimized code might restrict this
even more. ROOT strives to support as many platforms and operating
systems as reasonably possible. It certainly intends to support all
architectures and operating systems that are in production use, by
experiments or a significant number of physicists. It currently supports
GCC and clang compilers, up to the latest released versions; work is
ongoing to also support the Microsoft Visual Studio Compiler (0.2FTE/y).
The supported operating systems are Linux (Fedora, CentOS, Ubuntu,
Debian) and macOS (up to the latest version, including the ARM M1
version); work is ongoing to support Windows. Architectures include
Intel and AMD 64bit, Intel 32bit, ARM Aarch64; work is ongoing to ensure
continued support of Power8 and Power9. ROOT is released by Linux
distributions (or rather their package maintainers) on Fedora and CentOS
EPEL, through Conda, Homebrew, and Snap. This service has been provided
by the community for many years, and seems to demonstrate that ROOT is
part of a solid, sustainable software ecosystem. To facilitate all of
this, ROOT maintains a Continuous Integration and Testing / Continuous
Release system (0.5\,FTE/y).

In the past, being multi-platform caused ROOT to contain
architecture-specific code, such as X11 for Linux; Cocoa (including
ObjectiveC code) for macOS, and GDK for Windows. This is expertise not
needed elsewhere in ROOT, and the evolution of these platforms (such as
macOS dropping X11 and GL support) forces ROOT to invest in these areas
of (for ROOT) niche expertise. To address this, ROOT is developing
HTML-based web-GUI and web-graphics systems, to succeed and eventually
replace the architecture-specific GUI and graphics backends.

This was again tested and validated by Apple's recent move to its
ARM-based M1 architecture: within a few days, ROOT was passing most of
its own, extensive test suite (\textgreater600'000 lines of code) on M1.
In that respect, ROOT's Achilles' heel is the set of platforms supported
by llvm and clang. It is reasonable to expect that clang, one of the
major C++ compilers and tooling ecosystems, will always support HEP's
main architectures. So far this has always been the case; for instance,
Power9 support and Apple ARM M1 support appeared in clang/llvm before
they were needed by ROOT. Nonetheless, ROOT's extensive and special use
requires expertise to adapt to these new architectures and their support
in clang/llvm.

\begin{table}[h!]
\centering
\begin{tabular}{ |p{7cm}|p{7cm}|}
\hline
\textbf{Supported operating system} & \textbf{Supported
architectures} \\
\hline
Fedora, CentOS, Ubuntu, Debian & x86\_64, i686, Aarch64, Power8,
Power9 \\
macOS & x86\_64, ARM M1 \\
Windows 32bit, 64bit (work in progress) & i686, x86\_64 \\
\hline
\end{tabular}
\caption{Operating systems and their architectures as supported by ROOT.}
\label{tableOS}
\end{table}

\subsection{Code Reuse and Preservation}

Reuse of analysis code is common practice in HEP. Being a
community-agreed standard tool with a strong focus on backward
compatibility, ROOT helps with making code reusable. Frameworks that
attempt to formalize analyses, claiming to ease analysis reuse and
preservation, often increase the effective analysis complexity: they add
multiple ingredients to define an abstract analysis on top of a wealth
of libraries and software tools. ROOT prefers to reduce complexity,
thereby guaranteeing evolution and support of future architectures.

The platform independence and backwards compatibility of ROOT's data
format helps analysis preservation: old ROOT files can be read without
issues by the most recent ROOT versions, even on the most recent chips
such as Apple M1. By supporting as many platforms as possible, ROOT aims
to help with preservation: new platforms will likely be similar to what
ROOT already supports; migration should thus be feasible.

Being able to run ROOT on a future platform is not sufficient for
preservation and longevity. The community must also retain sufficient
technical expertise to address future challenges, such as for reading
ROOT files into a new memory layout. The ROOT project's core team of
long term contributors is essential in this respect.

The move towards event data models (EDMs) that can be read without
accompanying experiment libraries is helping with preservation: this
data can be read with ROOT alone, across decades. This approach has been
beneficial to HEP already since PAW and ZEBRA - file formats that can be
read and processed still today.

ROOT is engaging in discussions on a \href{https://indico.cern.ch/event/896473/}{common likelihood interchange
format for HEP}.
On one hand, such a format is a useful and effective documentation
feature, for instance for publications. On the other hand, it allows for
optimized implementations of likelihood functions outside of the highly
general RooFit framework, which inspires the continuous development and
optimization of RooFit itself.

ROOT believes that analysis frameworks will continue to thrive: ROOT's
role is to provide the building blocks that enable physics groups to
tailor mini-frameworks to their use. ROOT is continuing to ingest common
functionality where it is of general use to the community. This is
especially called for where ROOT can implement these tools in a better
way, for instance regarding performance, maintainability, or
accessibility. Examples only from the RooFit context include the
RooCrystalBall and RooJohnson probability density functions; RooStats
and HistFactory; ongoing efforts to integrate RooFit extensions
developed within ATLAS; CMS's and ATLAS's higher-level likelihood
building tools; and RooUnfold.

\subsection{Open Science}

ROOT participates in CERN's Open Science Policy Working Group. This is
only the most recent example of a long history of ROOT engaging with
open science and open data.

ROOT is engaged in the experiments' open data efforts, consulting on
issues with their data and analyses efficiencies (0.1\,FTE/y). ROOT also
benefits itself tremendously from said open data efforts: these open
data samples play a crucial role for ROOT's training sessions, example
code, benchmarks and tests. One of the limiting factors appears to be
the available personpower from the experiments on the open data side,
and ROOT would be happy to intensify its involvement.

All of ROOT's code is public. All changes to ROOT's code are public;
they can be reviewed by anyone in the world. ROOT's website is providing
everything from an introduction to ROOT up to the technical
documentation of all of its interfaces. ROOT sees contributions from
high school students to university professors, with changes in 2020 in
ROOT itself that correspond to a diff file of 31MB, with about 2800
lines changed each working day. ROOT follows the \href{https://fair-software.nl/}{FAIR-Software}
approach, making it as open as best practices recommend.

Multiple R\&D projects are taking place in the context of ROOT, for
instance with the \href{https://www.hzdr.de/db/Cms?pNid=2097}{LLAMA / Alpaka} group of the Helmholtz Center Dresden,
Germany.
While CERN is providing the backbone of ROOT's resources, crucial
long-term contributions for specific areas come from Fermilab, GSI,
Princeton, UCSD, and University of Nebraska. Additional temporary
commitments happen on a regular basis, in 2020 for instance by LAL,
through Google Summer of Code as well as Season of Docs.

ROOT's developments have an impact on other fields. A good example is
ROOT's interpreter cling, which is famously serving as the engine behind
\href{https://github.com/jupyter-xeus/xeus-cling}{Jupyter's C++ kernel}
and whose \href{https://reviews.llvm.org/D96033}{integration into llvm} is ongoing;
or ROOT's work on an improved, platform-independent RANLUX
implementation \cite{RANLUXPP}
that we expect to integrate into GCC's standard library and the GNU
Scientific Library. ROOT has strong ties with the C++ committee;
experience from HEP's use of C++ frequently leave traces in
modifications of C++ \texttt{std::simd} \cite{STDSIMD},
\texttt{std::variant} 
\cite{STDVARIANT},
C++ reflection \cite{CXXREFL,CXXREFLTS}
and -- due to the well-established connections -- in the C++
implementations such as \href{https://www.mail-archive.com/gcc-patches@gcc.gnu.org/msg261861.html}{\texttt{std::any}} and \href{https://gcc.gnu.org/bugzilla/show\_bug.cgi?id=87520}{\texttt{std::unique\_ptr}}.
ROOT's I/O format for the HL-LHC, RNTuple, will be accessible through a
library that does not rely on ROOT or its interpreter, with limited
functionality. Nonetheless we expect that this approach makes ROOT's I/O
layer significantly more interesting for other sciences.

This means ROOT is itself an active element of open science. At the same
time, ROOT is used also outside HEP, for instance through BioDynamo \cite{BioDynamo} or in
quantitative finance research \cite{HighLOVis}.

\subsection{Training, Education, Support}

ROOT's advances will only have an impact on the community if the
community is aware of them. Dissemination is a core responsibility of
ROOT; creating training material and presentations for new features is
time consuming. At the same time it is extremely beneficial also for the
evolution of ROOT's new interfaces, for instance through feedback
received during training sessions. In general there is a constant
tension between investing in ROOT's evolution, or investing in talking
about it; a suitable balance has to be kept (1\,FTE/y).

The ROOT team has started a "Train the Trainer" series of events which
will be resumed after the COVID-related travel bans are reduced. Its
purpose is to scale training out: if the ROOT team cannot train the
community the way it should, then the ROOT team should train
\emph{trainers} who will then multiply the effectiveness of ROOT's
training material. These "ambassadors" are community connections,
expected to give multiple trainings per year, and collect feedback on
ROOT and the training material, to commonly advance it. Keeping the
material with the project allows ROOT to evolve the material together
with ROOT, addressing new demands quickly and sharing new features
early. Nobody knows ROOT as well as its developers; both the community
and the ROOT project benefit from these close ties to the trainers and
the trained communities. Even once the ambassadors take over, ROOT
expects to continue to give introductory ROOT courses, for instance to
summer students (traditionally three per year) and as part of the
experiments' introductory courses for PhD students.

ROOT has engaged with the Masterclasses project, for instance \href{https://alice-masterclass.web.cern.ch/MasterClassInstallation.html}{with ALICE}
and tries to support them both on a technical level and by making HEP
more visible outside the CERN member states: following ROOT's workshop
in Sarajevo, including a public lecture to local students, the Sarajevo
University joined the Masterclasses program.

\begin{figure}
\includegraphics[page=1, width=1\textwidth]{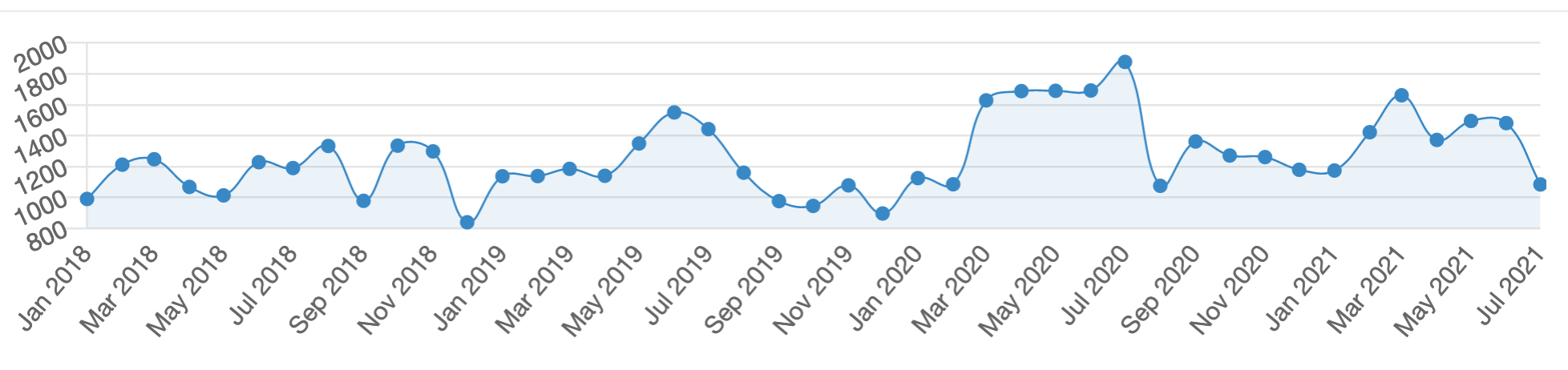}
\caption{Distribution of monthly ROOT forum postings in recent history.}
\label{ForumPosts}
\includegraphics[page=1, width=1\textwidth]{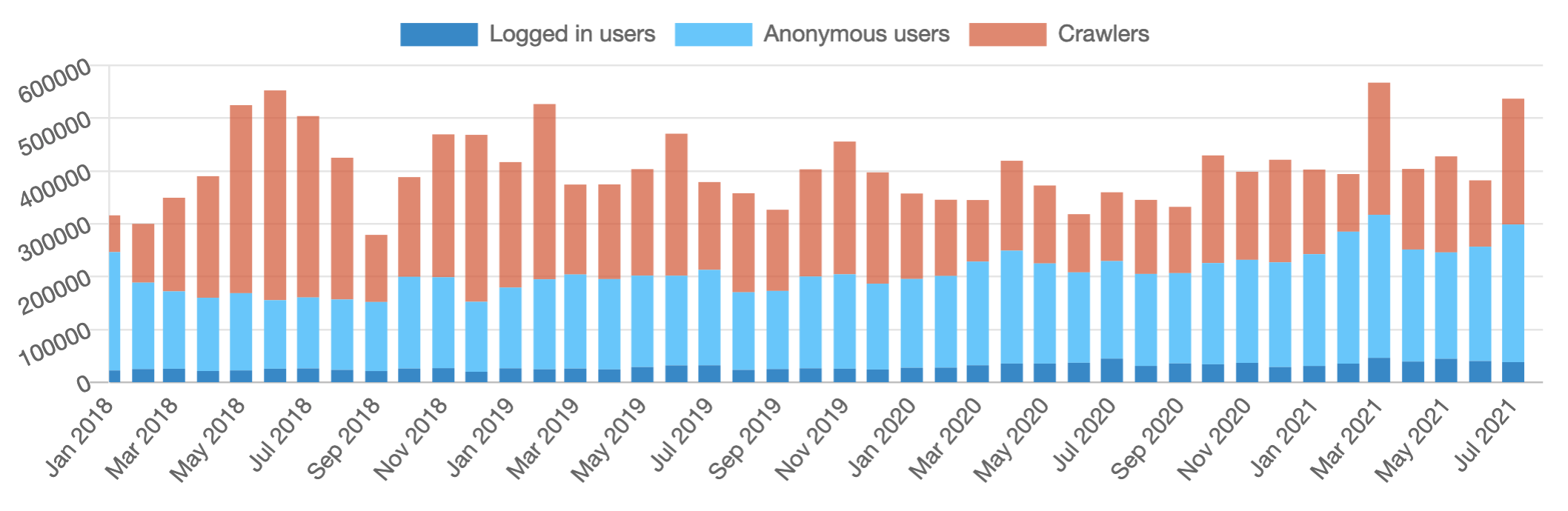}
\caption{Distribution of consolidated ROOT forum pageviews in recent history.}
\label{ForumPageViews}
\end{figure}

The ROOT forum sees vivid interaction from young physicists to senior
statistics experts, with monthly numbers that are usually significantly
higher than 300 thousand page views and thousand messages posted (see Fig.~\ref{ForumPosts} and \ref{ForumPageViews}).
Over
the last 12 months, the average response time to questions is below one
hour (3\,FTE/y shared among the project members). The forum is a
knowledge base for ROOT and HEP analysis in general, and a fantastic
indicator of where the community's problems are, and where the ROOT
project needs to invest.

With the creation of a new generation of ROOT interfaces, the ROOT team
expects to asymptotically reduce the investment in support. Robust,
simple interfaces with good documentation, excellent tutorials, and good
defaults deploying expert optimizations "behind the scenes" are expected
to reduce boilerplate code that gets "inherited" over generations of PhD
students. This will result in analysis code that is simpler to
understand, more robust, and can serve as a better starting point for
future PhD generations.

\section{Analysis Data}

\subsection{Data Models Optimized for Analysis}

Today's CPUs and GPUs, such as available through exascale systems,
benefit greatly from large amounts of data, allocated consecutively in
memory, that an algorithm can iterate over. This favors "structs of
arrays" (SoA) rather than the more traditional object-oriented "arrays
of structs" (AoS). Modern ROOT features enable SoA as input, handle
memory as SoA internally, and pass memory as SoA to GPUs.

This is further facilitated by ROOT's I/O format: TTree and RNTuple
store data in a columnar way. Reading this into memory provides, by
default, SoA layout. With TTree this was a lucky coincidence that became
relevant only after the interface design of TTree. RNTuple on the other
hand can immediately benefit from this, and makes columnar data access
as SoA a first-class interface.

While for many programs, AoS-to-SoA transformation is a costly
operation, HEP does it as part of ROOT I/O serialization. This means HEP
and ROOT have decades of experience with it, with the experiments'
analysis data formats migrating to columns (TTree "branches") of simple
data types. These "simple aods / ntuples" simplify data discovery and
reading, as they can rely purely on ROOT without the need for extra
libraries, reading data into the beneficial SoA in-memory layout.

SoA layout of simple data types has the additional advantage of being
language agnostic (or compatible): virtually all languages can handle
arrays of ints or doubles. This improves very efficient language
interoperability. The language boundary is crossed not once per value,
but once per array, making interoperability significantly faster as can
be seen by \href{https://github.com/root-project/rootbench/}{ROOT's benchmarks} with handing data into NumPy, machine
learning libraries, or GPU processing.

\subsection{Machine Learning}

As described in section~\ref{ch:AE-ML}, ROOT plays
a role as a data source for training and inference of machine learning
models. Currently, ML (whether run on GPU or not) benefits from large
arrays as input data. These data "structures" are simple enough that
ROOT - and here especially RNTuple - will be able to provide them with
virtually no overhead. With the advent of graph neural networks, data
must be "reformatted" into more complex memory data structures. Most
state-of-the-art ML algorithms are fairly young, also in relation to
when HL-LHC is supposed to start. The whole sector of quantum-computing
inspired ML has not seen production use yet. Given these expected
fundamental changes, it is essential for ROOT to have the capability to
adjust in-memory data structures, more than to optimize to current
requirements. Here, LLAMA can be a fundamental ingredient, together with
RNTuple's ability to handle custom data layouts and ROOT's C++
interpreter and just-in-time compiler (JIT) cling enabling optimized
transformation of memory layouts.

ROOT is expected to contribute to ML advances indirectly, with its C++
interpreter and JIT: As an example, trained models can be read and
converted to C++ code at runtime, with state-of-the-art optimizers
(provided by the compiler technology community) significantly
accelerating and simplifying inference \cite{SOFIE}.
Similar technology advances can be expected also in the future.
Leveraging those, ROOT can increase its visibility and relevance also
outside HEP, and provide benefits to HEP with little additional
investment, by building on ROOT's existing core components.

Many of these tasks require expertise in ROOT's key enabling
technologies that are part of and used by ROOT, such as clang, ROOT's
Python binding, or the I/O system. They also require an excellent
understanding of the ML application side and its requirements, as well
as credibility and community trust. This enables conception, design and
implementation of solutions with sufficient lead time ahead of demand,
and gradual ("agile") adoption. Given the required expertise and
development effort, ROOT can significantly impact efficient use of
compute resources, including GPUs, for analyses and other workflows
employing ML (reconstruction, simulation). This happens directly through
software technology advances relevant to HEP using ROOT's technologies,
and by encouraging and guiding the use of efficient analysis designs.

RDataFrame-based training on GPUs is completed by 2023 and ML-optimized
RNTuple-to-GPU transfer by 2025. Much of the evolution cannot be planned
ahead and needs to adapt to the evolving requirements of physics
analyses and ML research. The inherent risk is the lack of
predictability of ML's future evolution. We hope that a versatile setup
like ROOT's (and here especially that of RNTuple and RDataFrame) allows
us to adjust quickly to ML's upcoming requirements. Lack of delivering
said adjustments in a timely manner will drive ML research, and as a
consequence also analyses involving state-of-the-art ML, into other
ecosystems, causing segregation by use of different tools that are
potentially less suited, optimized, or adapted for HEP.

\subsection{Data Movement to GPU}

GPUs are ideal for instance for ML training; ROOT is working to
accelerate data movement from storage to GPUs. R\&D areas include memory
layout particularly suitable for GPU algorithms \cite{LLAMA}; \href{https://developer.nvidia.com/blog/gpudirect-storage/}{direct transfer} from storage to GPU, bypassing of
CPU;
use of \href{https://developer.nvidia.com/blog/optimizing-data-transfer-using-lossless-compression-with-nvcomp/}{compression algorithms} optimized for GPUs;
and total throughput optimization of these different options, possibly
combining them. Much of this sees very recent and ongoing technology
evolution such as
\href{https://github.com/NVIDIA/nvcomp}{nvCOMP} and
\href{https://searchstorage.techtarget.com/news/252503292/Nvidia-finalizes-GPUDirect-Storage-10-to-accelerate-AI-HPC}{DirectStorage};
ROOT is following these developments, making sure that they can be
captured for production use in the context of RNTuple, ROOT's future I/O
library. Here, RNTuple is expected to initiate and schedule the data
transfer to GPUs, making use of its abilities to provide high-bandwidth
I/O in a flexible way including, most importantly here, its scheduling.

As this is a highly evolving area, risks are mostly associated with a
design that prevents HEP from benefiting from these performance
improvements. Even after considering current R\&D in the design of
RNTuple, residual risks are associated with much of the GPU ecosystem
being closed source. As this is in the high performance area of
computing, ROOT expects that C++ will continue to play an important
role. This should allow ROOT to benefit from its C++ core, possibly
after adapting RNTuple, its supported compression algorithms, and its
scheduler to requirements of future high-bandwidth storage-to-GPU
interfaces.

\subsection{Data Location, Transfer, Caching}

On the other extreme of data transfer is remote I/O, where data is
transmitted from a data lake or any other non-local, medium to high
latency storage service. This plays a role in centralized storage,
possibly separating storage and computer centers geographically, as well
as in (on- and off-premises) cloud computing.

ROOT's approach is to make use of application-side knowledge to drive
I/O before the data is needed: after an initial latency cost, subsequent
data should arrive before it is needed by the processing code. This was
already exercised by TTree's asynchronous prefetching which was not
commonly used in production, likely due to insufficient investment in
the feature's robustness. With asynchronous I/O being at the heart of
RNTuple's design, this is supported from the get-go and the default mode
of operation.

To reduce the impact of remote I/O latency, ROOT combines transfer
requests across an entry range. This was exercised with TTree's
TTreeCache for many years, enabled by default since 2015. RNTuple
utilizes a similar mechanism for grouping I/O requests; its performance
is being tuned in the context of HPC file systems such as \href{https://indico.cern.ch/event/1009424/contributions/4246120/attachments/2204442/3748030/Benchmarking\%20HPC\%20-\%20openlab\%20technical\%20workshop\%2021.pdf}{Ceph/Lustre}
and remote I/O through Davix. Remote I/O through Xrootd is expected to
be implemented by 2023, in close cooperation with CERN's Xrootd
developers.

Determining which columns need to be read can be based on past usage or
can be configured by the application. With efficient "read what you
need" and asynchronous I/O, pre-placement of jobs and data files should
be superseded by ROOT reading only those bytes that the application
needs, ahead of time, reducing I/O requirements and increasing CPU
efficiency.

ROOT allows for gradual production of derived data columns, such as
calibrated jet pTs derived from uncalibrated jet pTs, among others, and
extending an original dataset. These additional columns can be stored in
dedicated files, extending the original tree ("friend trees"). This
together with columns existing only for certain entries allows a storage
efficient creation of sub-samples, where overlaps between samples are
not duplicated.

Analyses see repeated runs on identical input data. ROOT is working on a
transparent, multi-tier caching layer to be deployed on batch systems
such as Spark clusters: ROOT's upcoming Distributed RDataFrame will
favor the "sticky" distribution of jobs' input data to enable node-local
data caches. This has been shown to accelerate processing \cite{RDFCache}.
Given sufficient resources, ROOT hopes to extend this
to an automatic data cache of intermediary analysis results, to
short-cut and dramatically accelerate the re-running of analyses.
Such a mechanism is considered for deployment in analysis facilities (\href{https://iris-hep.org/projects/servicex}{"ServiceX"}) that come
with their very distinct software ecosystems.

Given the relevance of object stores in next-generation data centers (such as \href{https://www.alcf.anl.gov/aurora}{Aurora}) as
an alternative to conventional filesystems,
there are ongoing efforts to support alternative backends in RNTuple,
such as Intel DAOS \cite{RNTupleObj}
or Amazon S3.
The use of object stores as a cache for accessed data is
also being investigated \cite{RDFObjCache}.

The risks associated with remote I/O are suboptimal usage of network and
CPU resources - corresponding to the key resources of HEP computing. For
the ROOT project it is of paramount importance that its I/O layer is
capable of making adequate use of these resources; benchmarks are
TTree's behavior as well as raw network speed and CPU usage. Given the
pace of technology evolution in the network and remote storage
environments, ROOT does not expect technology evolution to be a
significant risk on the HL-LHC timeframe.

"Black-box" systems optimized for re-processing of analyses can be a
motivation for analysis physicists to migrate to alternative ecosystems.
This is a residual risk that ROOT tries to reduce through technology
advances and acceleration of repeated analysis. From ROOT's experience,
these optimizations are rarely applicable to multiple analyses, for
instance on the level of an analysis facility: reuse of data and
especially intermediary analysis results are expected to be specific for
a given analysis.

\section{Analysis Design}

The "old" ROOT forced physicists to deal with the data source, reading,
and the event loop. This prevented many optimizations, for instance
multithreading. PROOF tried to formalize analyses, enabling ROOT to
process the analysis concurrently. Still, for optimal efficiency, the
analysis was commonly interacting directly with the I/O layer, and
handling the values read from storage, and their types, explicitly.

In 2018, RDataFrame became ROOT's modern analysis interface,
revolutionizing how physicists write analyses today. As can be seen from
ROOT's user forum, RDataFrame is now a topic as popular as "tree" or
"histogram". This can be seen as a recent major success and a crucial
contribution that the community seems to have been longing for. The
significance for the community is certainly similar to RooFit.

Unlike alternative approaches, ROOT believes that analyses should be
allowed to be conceived and written "event-centric": the minimum number
of jets per event, maximum missing ET, number of jets associated with a
muon - all these parameters are traditionally per event. ROOT expected
it to be much easier for physicists to not give this up when writing
their analysis, but instead to handle arrays of events (and arrays of
arrays of jets and muons) "behind the scenes", where concurrent and
batch processing can run on the user declared analysis, without
physicists having to think about the array-ness of certain values such
as jet pT.

\subsection{Declarative Analysis}

RDataFrame's main benefit is its declarative interface style, where
physicists declare \emph{what} needs to be done, allowing ROOT to take
care of \emph{how} to do this optimally. As an example, to consider only
events with more than two jets reads Filter("njets \textgreater{} 2").
This is done with minimal overhead, yielding a very efficient,
multi-threaded analysis, which evaluates a complex analysis graph in a
single pass through all input data.

The community has built several "mini-frameworks" on top of
RDataFrame, for instance \href{https://indico.cern.ch/event/948465/contributions/4324161/}{\emph{bamboo}}, 
\href{https://github.com/KIT-CMS/CROWN}{\emph{CROWN}},
and \href{https://github.com/lcorcodilos/TIMBER}{\emph{TIMBER}}. This
by itself is a good sign, showing that ROOT has successfully provided
another fundamental building block significant enough to be picked up.
ROOT is following closely what these frameworks provide, and what they
have difficulties in providing, to "fill the gaps" and - where it is
adequate - to provide centrally maintained facilities. Examples of
current developments include the variation of event weights and other
analysis parameters, to determine the effect of uncertainties. This will
be possible without re-running the whole analysis for each variation.
This, too, is expected to have significant effects on how HL-LHC
analyses will be written.

While RDataFrame supports multithreaded analyses from the start,
multi-node support is introduced by Distributed RDataFrame. The latter
works together with schedulers and job submission backends (Spark, Dask,
etc) for an optimal distribution of input data and compute tasks to
clusters of machines. Such clusters are often institute clusters,
benefiting from a much higher data-reuse and code-rerun rate than for
instance the grid. This motivates ROOT's work on backend-specific work
placement, which will allow for analysis acceleration through node-local
caches. In order to enable support for a variety of use cases,
distributed RDataFrame features a modular backend design. In the future,
users will be able to distribute the same computation graph over a set
of different cluster frameworks by changing a single line of code.
Usability is a key ingredient for such a distributed analysis feature,
and configuration of underlying scheduler backends, authentication, data
placement, and writable disk space and result collection is generally
complex. ROOT plans to address this with a community-wide,
community-maintained database of available configurations, such that
physicists can use Distributed RDataFrame with their local resources
simply by providing the cluster identification.

As with any new ROOT interface, RDataFrame is written with Python in
mind; Distributed RDataFrame is only possible because of this. Several
optimizations are pending, for instance for optimized just-in-time
compile code; compiled code generated from Python code to be run in the
"hot" event loop; more efficient bulk processing. Much of this requires
understanding of RDataFrame, ROOT's C++ interpreter and just-in-time
compiler called cling, PyROOT, and Python in general.

\begin{minipage}{0.9\linewidth}
\begin{lstlisting}[language=C++,numbers=none,caption=RDataFrame C++ code example.]
ROOT::EnableImplicitMT();
ROOT::RDataFrame df(dataset);
auto df2 = df.Filter("x > 0")
             .Define("r2", "x*x + y*y");

auto rHist = df2.Histo1D("r2");
df2.Snapshot("newtree", "out.root");
\end{lstlisting}
\end{minipage}

\begin{minipage}{0.9\linewidth}
\begin{lstlisting}[language=Python,numbers=none,caption=RDataFrame Python code example.]
ROOT.EnableImplicitMT()
df = ROOT.RDataFrame(dataset)
df2 = df.Filter("x > 0")
        .Define("r2", "x*x + y*y")

rHist = df2.Histo1D("r2")
df2.Snapshot("newtree", "out.root")
\end{lstlisting}
\end{minipage}

As can be seen from the uncertainty variation of parameters, investment
can have a significant performance impact for analyses (O(10) speed-up
for regular analysis, compared to TTree, with an investment of 0.5\,FTE
for 2 years), running the analysis once instead of once per variation.
ROOT plans to invest in RDataFrame to make it significantly easier to
use with ML, RooFit, and GPUs (O(10) speed-up for average analysis, with
an investment of 2\,FTE for 4 years). Together with snapshotting of
intermediary results, optimizations of the Python and just-in-time
interfaces, and internal bulk processing of RNTuple data, common
analyses will be accelerated by at least an order of magnitude. We
expect that by the time of HL-LHC, the vast majority of analyses will be
using RDataFrame - especially as already now, RDataFrame is as popular
on ROOT's user forum as ROOT's histograms or TTree. Such investment thus
scales out to a very large number of analyses, with a significant
community impact. It also fosters ecosystems to be built around
RDataFrame - tools that can be shared and integrated into ROOT if they
are of general use.

This acceleration can bring analyses that are too time consuming into
reach; it can improve analyses' ability to optimize their parameters.
This is beneficial for traditional analyses with iterative analysis
optimization, and a prerequisite for differentiable analysis \cite{DiffAnalysis}
approaches.

\subsection{Accelerators}

At the HL-LHC timescale, ROOT expects GPUs to be the commonly available
accelerators. For analysis, their main use will certainly continue to be
ML training. Also because of the expected ubiquitous use of ML for
HL-LHC analyses, ROOT expects GPUs to be ubiquitous for analyses, and is
working on R\&D to integrate GPU acceleration in RooFit which builds
upon prior work on architecture-specific vectorization \cite{FastRooFit}.
This shows very
promising results. To be of general use, such acceleration must be
enabled by default, when available and advantageous. Much of the R\&D
goes into determining mechanisms that transparently turn on such
acceleration.

ROOT's future I/O subsystem RNTuple is designed to utilize
(de-)compression accelerators, which can play a crucial role in analysis
throughput. These have already been available in past Xeon generations.

\section{Analysis Algorithms}

\subsection{Histogramming}

ROOT's histogramming package has been the core ingredient for most analyses,
for decades. Hardware performance characteristics have changed since its
original conception, and much of the design is limiting today's and
tomorrow's use. The interfaces of ROOT's histograms are not designed for
their usage from Python, nor their interoperability with the Python
ecosystem such as NumPy. Alternatives, such as \href{https://github.com/boostorg/histogram}{Boost histograms},
have been created. They lack commodity features expected from a
ROOT-provided histogram package, and are optimized more towards raw
performance than usability. ROOT's histograms must provide graphics,
serialization, and fitting facilities.

The creation of alternative packages and the "impedance mismatch" of
modern code with ROOT's histograms (including ownership, lack of
generalization, more than 300 member functions, no support for atomic
bins) clearly signal that this code part needs to be revisited. Thanks
to a partial FTE contribution from LAL, ROOT made much progress with a
new \href{https://github.com/root-project/root/blob/master/tutorials/v7/simple.cxx}{histogram library}.
Nonetheless, without sustained effort for two years, a new ROOT
histogram library cannot be advanced to the minimal feature level
required for adoption. The community risks continued work around the
ownership discrepancies of ROOT's histograms, paying a performance price
for analyses, and additional fragmentation in the experiments' online
and offline framework, where even today, experiment specific
histogramming facilities have been reintroduced to not use ROOT's
histogram library \cite{DQM}.
CMS has provided a \href{https://indico.cern.ch/event/697389/contributions/3108820/attachments/1714101/2764583/cms-dqm-rootworkshop.pdf}{review}
of issues of ROOT histograms for DQM. Despite these issues, ROOT
histograms currently remain the most commonly used histogramming
library. Without investment, usage will further segregate, with no
obvious place to invest for the whole community.

This new generation of histograms will allow ROOT to implement long
requested features, such as concurrent filling, consistent availability
of arbitrary dimensions, or special axes such as circular axes or
logarithmic axes with a custom base, efficiency axes (pass versus all),
or axes for multiplicity studies. While some of these features merely
simplify writing an analysis, some other features counter commonly seen
correctness issues with histograms, such as floating point precision
issues with multiplicity axes, or can help reduce memory use in
concurrent analyses such as analysis trains.

\subsection{Modeling and Fitting}

RooFit is high energy physics's standard tool for modeling statistical
distributions and building likelihood functions. It is used in most LHC
analyses for estimating physical parameters, confidence intervals and
discovery significances. The minimization of a likelihood function
defined in the RooFit framework is central in most LHC analyses. With
the HL-LHC, the number of parameters and observables in the likelihood
function is expected to increase. Usually, minimization time grows
superlinear with the number of parameters, meaning technical innovation
in likelihood minimization is necessary.

Thanks to modern deep-learning
frameworks such as TensorFlow or PyTorch, it is increasingly easy to
minimize a function expressible as a chain of linear algebra operations.
However, the mathematical operations in a typical LHC likelihood fit are
often much more complex, requiring for instance the numerical
integration of probability densities for normalization purposes or
looking up the result of an auxiliary measurement in a histogram. The
RooFit library was designed with such general likelihood functions in
mind, supporting unbinned minimization. As RooFit is written in C++,
there is still ample room for performance optimization at all the levels
of the likelihood evaluation and minimization, making it ready to face
the challenges of the HL-LHC data volume.

\begin{figure}[H]
\includegraphics[width=1\textwidth]{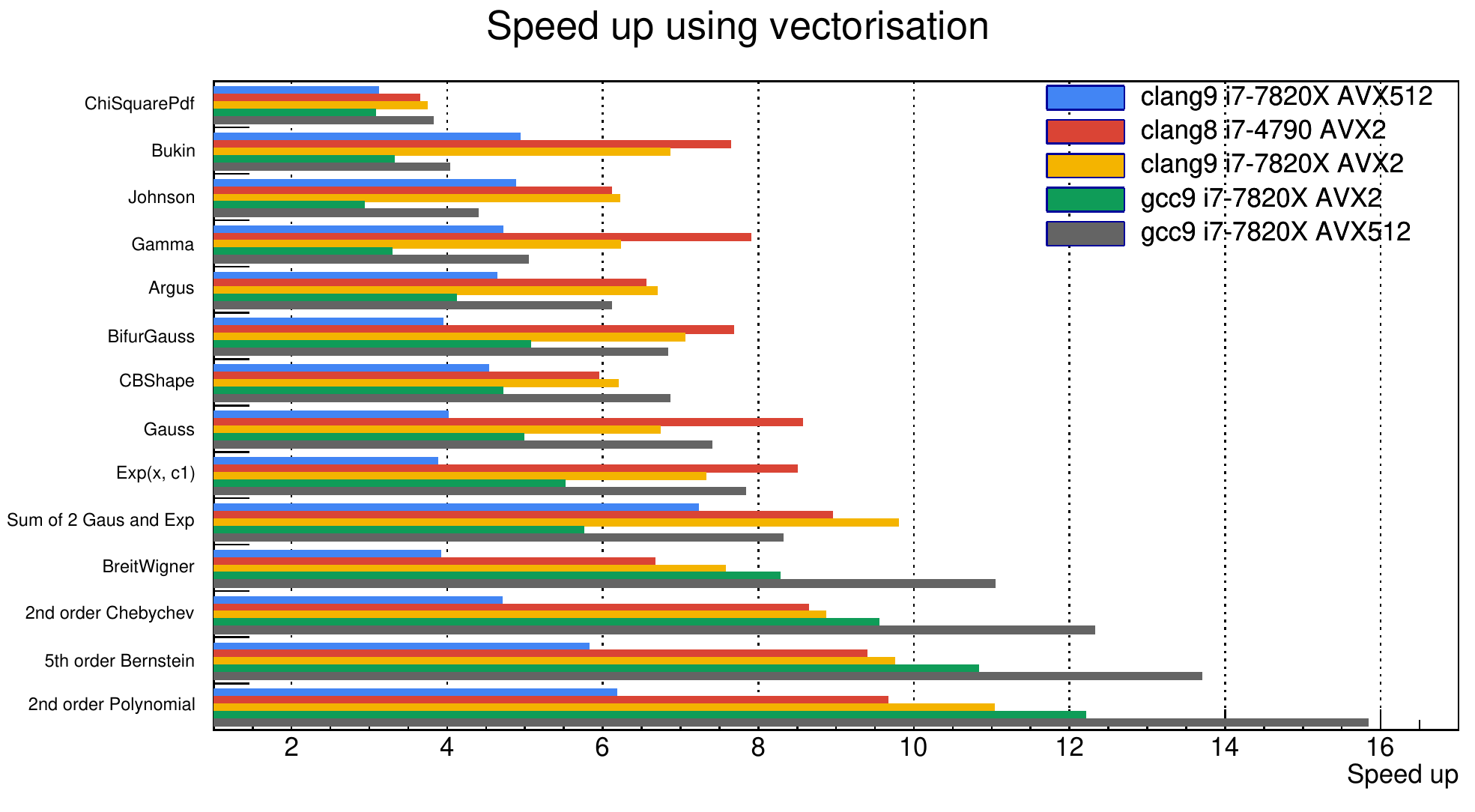}
\caption{Acceleration from vectorized functions in RooFit.}
\label{fig:RooFitSIMD}
\end{figure}

The ROOT developers pursue several paths to optimize RooFit's
performance: speeding up the likelihood gradient calculation,
accelerating specific computations with GPUs, writing more vectorizable
code (see Fig.~\ref{fig:RooFitSIMD}), general optimization of expensive operations, and improving the
interoperability with other libraries that can handle large datasets.
The larger the number of parameters in the likelihood function, the more
expensive it is to numerically determine the gradient by varying one
parameter at a time. In late 2021, RooFit will introduce functionality
to parallelize the gradient calculation over multiple CPU cores. In the
following year, R\&D on auto-differentiation for gradient computation in
constant time will move into the focus (see Minimization section). To
increase the throughput of computations, much of RooFit was recently
rewritten to support auto-vectorization. This path will also be followed
in the future, alongside the continuous optimization of CPU code in
general. In 2021, the foundational work to offload computations to a GPU
was done. Next, the remaining likelihood-building blocks that are
frequently used for HL-LHC analyses have to be implemented.

With the increasing dataset sizes, it is crucial to improve RooFit's
interoperability with other libraries meant to deal with big data. This
concerns other ROOT components (RDataFrame and RNTuple) and libraries
from the Python ecosystem (e.g. NumPy and Pandas).

A redesign of the RooFit interfaces based on value semantics instead of
RooFit's current reference semantics would make RooFit significantly
easier to use, simplify ownership management, and enable the same
straight-forward interfaces from C++ and Python. This will attract
physicists who would otherwise use Python-AST based interfaces with
reduced performance and feature characteristics.

\subsection{Minimization}

\begin{figure}[H]
\centering
\includegraphics[angle=90,width=0.7\textwidth]{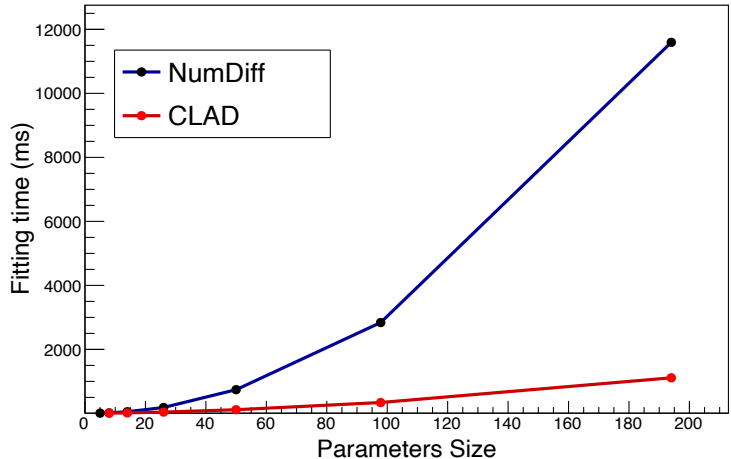}
\caption{Speed-up of multi-parameter minimization due to automatic differentiation.}
\label{fig:clad}
\end{figure}

Minimization is another area that can benefit greatly from the use of
GPUs. The evaluation of minimizer functions can be accelerated on GPUs,
especially when operating on large amounts of data. Use of analytical
gradients, or automatic gradient calculation for instance with \href{https://github.com/vgvassilev/clad}{clad}
can provide further acceleration, see Fig.~\ref{fig:clad}. Some of the algorithmic challenges
remain hard, such as normalization of probability density functions when
evaluating likelihood functions as seen with RooFit models; we are
confident that continued R\&D investment will pay off.

Advances in minimizers will provide immediate benefits to RooFit. Even
the side-effects from these R\&D efforts are useful to the community and
outside, with the C++ automatic differentiation tool clad being a
perfect example.

ROOT continues to try state-of-the-art minimizers, benchmarking them
against MINUIT and other production minimizers. Production-grade,
general purpose minimizers are rare; ROOT does not expect that major
development effort will need to be invested here, except for the
above-mentioned GPU minimization.

\section{Visualization and Graphical Interfaces}

ROOT's graphics style and abilities have defined the visual language of
HEP. With the ever increasing complexity of architecture specific
graphics and the web taking over GUI and graphics, ROOT has opted for a
redesign of its graphics and GUI systems based on HTML, CSS and
JavaScript.

\subsection{Scientific Graphics}

ROOT's graphics system allows extreme configurability while creating
graphics in publication quality. Several of the defaults would be chosen
differently today, for instance the font size being relative to the
window size, or the line widths of histograms, or histograms' default
colors.

The complexity of maintaining highly efficient graphics on all platforms
has increased enormously: Linux is migrating from X11 to Wayland (and
alternatives); macOS offers Cocoa and Metal; Windows offers GDI+ or
DirectX. Luckily, web browser engines have established themselves as a
platform independent abstraction layer. The Chrome / Chromium Embedded
Framework for instance is used by many programs such as Mattermost, the
Atom editor, and Spotify. This means ROOT's architecture specific
libraries can be succeeded by libraries maintained by the open source
community.

Any scientific graphics library is built around primitives such as axes,
lines, greek letters and formulae, and the ability to zoom. Luckily,
these features are available in open source JavaScript graphics
frameworks that ROOT was able to adopt and build upon for its usage.
JSROOT is ROOT's JavaScript interface that draws for instance histograms
in a browser window, with virtually the same configurability and
performance as the previous architecture-specific implementations.
Hundreds of histograms can be shown interactively, with smooth
interaction.

The move to web graphics enables embedding of ROOT graphics in custom
GUI applications such as Qt applications. Before, embedding of ROOT
graphics depended to a large extent on platform-specific features (and,
as a consequence, stopped working on macOS unless using legacy X11
implementations). Today, virtually all GUI systems allow embedding of
browser windows, or are written themselves as browser GUIs.

With this new technology in its hands, ROOT is now working on making the
user facing implementation of ROOT's graphics much simpler to use. New
interfaces allow ROOT to define new defaults and to streamline the
graphics interfaces, simplifying for instance ownership management and
separating data structure from graphics abilities. This work is ongoing
and expected to deliver a robust, simpler graphics interface in time for
HL-LHC. Prototyping new graphics programming models is expected to
conclude in 2022. Implementation of the new default style and CSS-based
style customization is expected to be available by 2024. Grid deployment
(for instance with the above mentioned Chrome Embedded Framework
available through LCG) is expected to be achieved by 2022.

\subsection{Graphical User Interfaces}

With the graphics system moving to web technology, moving ROOT's custom
GUI system along was an obvious next step. ROOT has decided to use the
feature-rich, open source Web GUI library named OpenUI5. Communication
of the browser window with the ROOT process happens through an open
source web server library called civetweb, which is developed
independently of ROOT. The ROOT process communicates through its
interpreter / JIT compiler cling, passing data both ways, and calling
the appropriate methods on the C++ or JavaScript side. ROOT's I/O
subsystem is used to transform C++ objects into JSON and back, to
transport state between the ROOT process and the browser.

This shows that ROOT is able to re-use its own core components to
generate a wealth of fundamental tools specific for HEP. While ROOT's
GUI itself is important (for instance ROOT's new RBrowser, or the new
fit panel), it is equally important that ROOT and its interactivity can
be embedded in "foreign" GUI systems, and retain its functionality and
interactivity. The current setup based on web technology allows just
that, as demonstrated for instance by the ROOT masterclasses or Eve7.

\subsection{Jupyter Notebooks}

For "exploratory analysis", i.e. the first steps into discovering data,
fundamental distributions, or even for training of machine learning
models and the evaluation of their quality, Jupyter has established
itself as one of the interfaces. Being based on web technology itself,
it integrates nicely with ROOT's web graphics already today,

ROOT's C++ interpreter / JIT compiler cling is the engine behind the
official C++ Jupyter kernel. ROOT provides an enhanced version of this,
making many ROOT features more accessible through its own, dedicated
kernel.

Work is ongoing to integrate ROOT's web GUI system with the new Jupyter
GUI system. Given sufficient demand from physicists, this feature is
expected to be available in 2024.

\subsection{Event Display}

\begin{figure}
\includegraphics[width=1\textwidth]{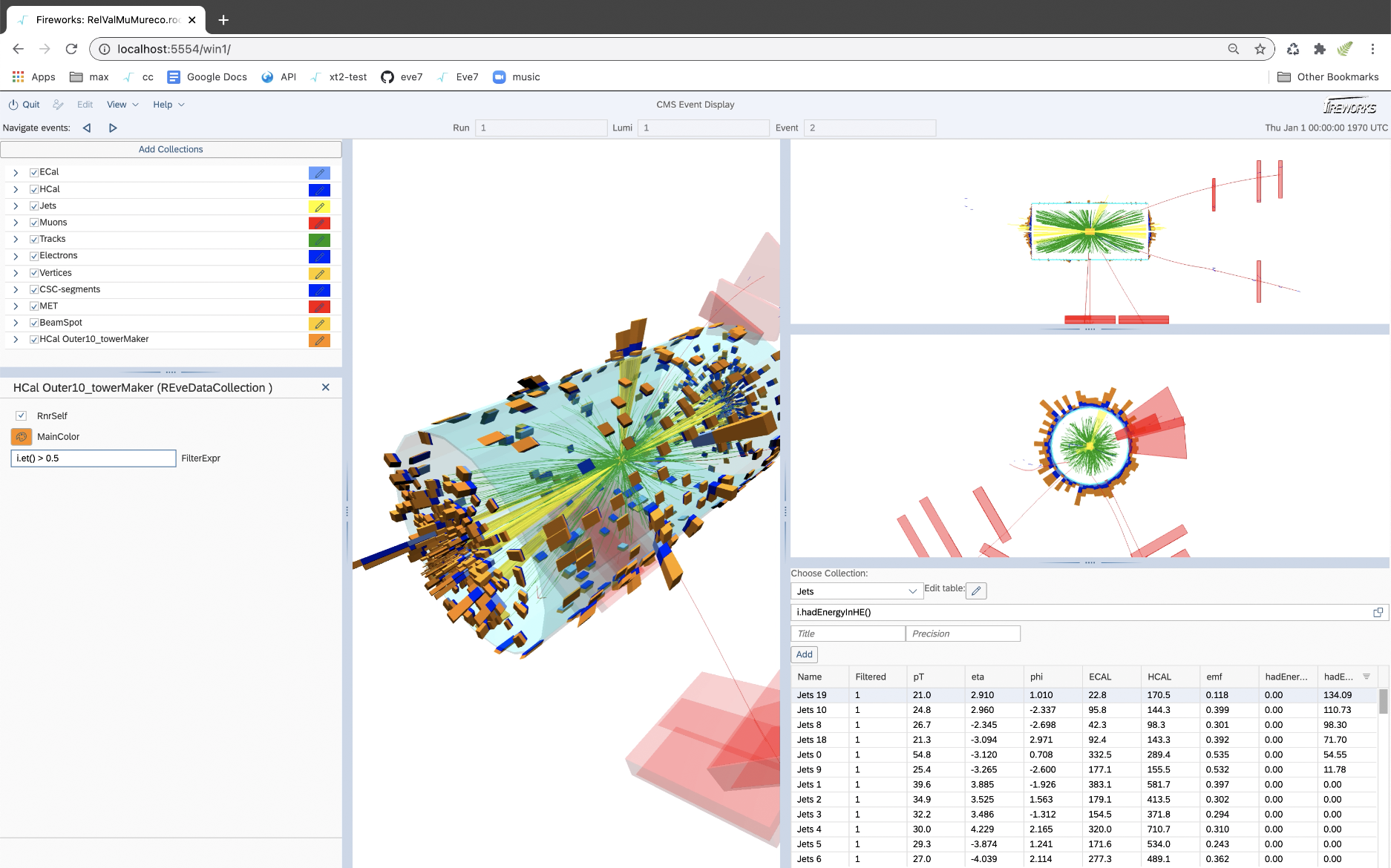}
\caption{ROOT event display with object selection.}
\label{fig:REve}
\end{figure}

ROOT's event display EVE was in production use in many experiments, for
instance CMS, ALICE, Belle II, T2K, HyperK, ILC, NA62 and several
smaller experiments in neutrino, nuclear, and medical physics. Being
exposed to the same problems and limitations as ROOT's legacy graphics
and GUI libraries, Eve has been redesigned as Eve7, to work with ROOT's
new, web-based graphics and GUI system, too; see Fig.~\ref{fig:REve}.
This has increased its
versatility dramatically: Eve7 works on all platforms, can present
events remotely, and is thus a perfect implementation for everything
from online monitoring to outreach.

Apart from JSROOT and ROOT's GUI system, Eve7 also uses ROOT's C++
interpreter / JIT compiler internally for filter expressions, for
instance to display only jets satisfying certain cuts. As these
expressions can use the full power of C++, the event display is highly
customizable and extremely versatile, useful for instance to
understanding effects of detectors or detector algorithms.

Event displays play a role in analyses, too: they allow a visual
inspection of selected events, for sanity-checking and to determine for
instance different topologies of samples. As such they are inherently an
analysis feature, where usability and performance is a key requirement.

Being developed in close collaboration with CMS and currently deployed
as a prototype in CMS, the feature set and performance of Eve7 has
already impressed several users of event displays. A first experiment,
Mu2e at Fermilab, has officially announced their interest in using Eve7
in production in the coming years. ROOT has every reason to expect
community adoption of Eve7 to be at least on par with the adoption of
the legacy Eve.

We expect that the majority of effort will be invested into Eve7
satisfying the requirements of CMS, which can serve at the same time as
a proof of being feature complete. This work is expected to conclude in
time for HL-LHC, and driven mostly by CMS itself.

\section{Project Requirements}

\subsection{Organization}

The ROOT project is in its core a group of contributors to a common open
source project. Decisions are taken publicly at a weekly team meeting,
upon discussion and by consensus. Consensus building happens through
good arguments; ROOT is thus in practice a meritocracy, where new,
capable contributors can quickly acquire significant influence.

ROOT development was traditionally split in several loosely coupled
parts - both regarding libraries and regarding development subgroups,
for instance for the I/O or the statistical part of ROOT. In recent
years, these delineations have reduced, to encourage both sharing of
expertise, and exposing of problems and solutions to a wider audience,
with a more diverse background of technical expertise.

ROOT continues to be a successful integrated "hub" for R\&D; several
grants were attributed to work in the context of ROOT. Examples include
the R\&D on the \\href{https://reviews.llvm.org/D96033}{integration} of ROOT's C++ interpreter / JIT compiler
cling into llvm / clang ("upstreaming"),
compression studies \cite{Compression}, PhD students' work on
topics such as \href{https://github.com/alpaka-group/llama}{LLAMA}
and data caching for distributed computing \cite{RDFCache}.
Some R\&D - such as the core RNTuple work, or
RDataFrame - requires sustainability that goes beyond the time frame of
PhD student or grant requests, and needs to be taken up by long-term
contributors.

Besides this R\&D aspect, ROOT has a strong responsibility for
sustainability and support. ROOT's interfaces need to be backward
compatible, to facilitate sharing of analysis code across ROOT versions.
This requires an "interface vision": designing interfaces now that allow
optimizations behind the scenes in ten or 20 years from now is an
expertise that is required by the core ROOT developers.

Support is a significant workload for ROOT's core developers, with
hundreds of messages needing to be processed per day; issue reports and
code contributions requiring attention; infrastructure work
(configuration, continuous integration and benchmarking systems); and
the ROOT team presenting ROOT's advances at physics conferences, for
instance \href{https://indico.cern.ch/event/868940/contributions/3814685/}{iCHEP 2020},
\href{https://indico.jlab.org/event/247/contributions/3063/}{EIC},
\href{https://indico.ihep.ac.cn/event/11444/session/12/contribution/170}{CEPC
2020},
\href{https://indico.desy.de/event/28202/contributions/105604}{EPS-HEP 2021}, and during training sessions. This load can
generally not be shouldered by contributors who have dedicated R\&D
goals, but is carried by those developers with long-term contracts.

ROOT has virtually constant interaction with the experiments, from early
discussions on ROOT's development plans to high-priority issue
processing. The ROOT team engages with the experiments' software and
computing experts at a dedicated ROOT / Experiments meeting series; at
the Librarian and Integrators Meeting; and at the Architects' Forum.
Office visits between core experiments' developers and ROOT are common
(or rather, were common before the pandemic), as was pair debugging.
Requirements for the experiments' core software is collected through the
discussion of the plan of work and through issues opened by the
experiments.

This constant interaction leaves traces with all parties involved. ROOT
has contributed to the definition of the ALICE analysis model of Run3,
to the interplay of ATLAS XAODs and RDataFrame, and to optimizing CMS
nanoaods. Multithreading bottlenecks are addressed in cooperation; I/O
performance improvements are guided by measurements from the
experiments, often with tools provided by ROOT. With for instance
RNTuple, ROOT further increases its investment in benchmarking campaigns
with the experiments, to guide design and development by realistic,
early feedback.

Requirements for the experiments' analyses are generally much harder to
define. ROOT invites to workshops (e.g.~in \href{https://cern.ch/root2018}{2018} and \href{https://root.cern/blog/workshop_2015/}{2015})
to solicit combined feedback, makes use of the forum and GitHub issues
as feedback platforms, and presents where physicists are (see the list
of recent contributions at physics conferences above, regular
invitations at experiments' meetings, engagement / participation for
introductory ROOT courses) to engage in discussions.

ROOT owes much of its stability to the experiments' investment: all
experiments test all main ROOT releases, or even their branches; some
even report with low latency (about one week) on issues found in ROOT's
main development branch ("master"). This allows ROOT to carry out major
changes, in a coordinated way with the experiments. Recent examples
include the migration to a new PyROOT implementation; or updates of
LLVM. Another example of ROOT's engagement with experiments is the
\href{https://github.com/cms-sw/cmssw/pull/33825}{integration of RNTuple in CMSSW},
a combined effort of CMS and ROOT. The experiments trust ROOT to
maintain Vc, VecCore, and VecMath, packages used by the experiments
directly or indirectly, for instance through Geant4.

\subsection{Development Team}

ROOT's development is driven by R\&D contributions, where short term
contracts dominate by number. ROOT currently hosts one technical
student, one PhD student paid by CERN; one PhD student funded
externally; 2.5\,FTE CERN fellows; one CERN EP R\&D fellow; and one
externally funded fellow, i.e. in total 7.5\,FTEs.

Sustainability, guidance, accretion and integration of expertise is
guaranteed by developers with long-term positions, totalling 8.25\,FTEs, see Table~\ref{table:contrib}.

\begin{table}[h!]
\centering
\begin{tabular}{ |p{1.5cm} p{9.5cm} p{2cm}| }
\hline
\textbf{FTE} & \textbf{Expertise} & \textbf{Funding source} \\
\hline
5.5, incl 4.5 ICs & I/O, statistics, ML, C++ interpreter / JIT
compiler cling, ROOT's type description system, PyROOT, build system,
documentation, platform support and continuous integration
infrastructure, support & CERN \\
0.5 & I/O & Fermilab \\
1 & R\&D and support: event display & UCSD \\
0.5 & R\&D: C++ interpreter / JIT compiler cling; C++ automatic
differentiation & Princeton \\
0.5 & R\&D: compression, I/O & UNLincoln \\
0.25 & R\&D: web-based GUI and graphics & GSI \\
\hline
\end{tabular}
\caption{Current long-term contributions to ROOT.}
\label{table:contrib}
\end{table}

A fair fraction of these developers (marked "R\&D" in above table)
cannot contribute to the project's baseline load of supporting users on
the ROOT forum, infrastructure work (such as maintenance of ROOT's
continuous integration system or build system), or issue processing.
Much of the recent ROOT developments and expertise associated with those
developments - notably RDataFrame and RooFit - rely on fellows. RNTuple
is an exception from this, where the main developer was rewarded with an
IC and expects to migrate to ROOT for the vast majority of his working
hours.

The age profile of ROOT's development team versus its recent massive
renovation is one issue; the lack of gender diversity is another. ROOT
has succeeded in capturing physicists \emph{and} computer scientists;
developers from diverse national backgrounds. But with currently zero
female team members paid by CERN it has failed to achieve a reasonable
gender balance. This is an issue that ROOT expects to address in the
coming years, by investing in recruitment, reaching out to universities
and job fairs. Improving this requires a noticeable effort from the ROOT
project. This effort can only be effective in a sustainable way as
appropriate positions become available.

\subsection{Sustainability}

For ROOT as a long-term software project, sustainability is a key
requirement. Apart from technical sustainability such as I/O and
interface backward compatibility, separation of stable user interfaces
from evolving optimizations behind the scenes, and multi-platform
support, ROOT needs to also guarantee sustainability on a non-technical,
"cultural" level. Worth mentioning are ROOT's long-term common vision
for a ROOT evolution that matches HEP's requirements. ROOT's expertise
is the basis for the community's trust in ROOT and its continuing
evolution. The successful and established mode of working of ROOT's
development team guarantees productivity and integration of many kinds
of contributions, sources of contributors, and an active, vivid, and
continuously challenging ecosystem with constant communication with
ROOT's stakeholders.

ROOT's significant investment in support is addressed with a new
generation of coherently designed, well integrated, robust interfaces
that are simple to use correctly, and difficult to use incorrectly. This
should asymptotically lower the support load for ROOT.

Over the past years, ROOT has deprecated and removed interfaces, and
even large parts of ROOT. They either have seen insufficient use to
warrant the project's continuous investment; or they have not seen
general adoption from the community, thus benefiting from a lifecycle
independent of ROOT. Some interfaces (such as TLorentzVector) see
continuing wide-spread adoption but are nonetheless seen as problematic;
the ROOT team is working through documentation and training to migrate
usage to recommended interfaces (such as PtEtaPhiMVector).

At the same time, ROOT targets development and investment in a very
focused way. With about 5 million source lines, ROOT needs to choose
what to evolve, and what to "freeze". Many parts of ROOT have existing
usage, nonetheless much of their motivation is not applicable anymore
today. Examples include ROOT's collection classes which were needed
because no general C++ standard library support existed 20 years ago; or
parts written to bridge between ROOT's previous, limited C++ interpreter
and non-ROOT libraries; or internal tools that have been replaced by
other, open-source tools such as the transition from THtml to doxygen,
or TThread to tbb. Much of this legacy code has been written with a past
understanding of "software development's best practices", making any
more fundamental investment a challenge and of dubious value. Instead of
causing friction and additional work for the community, ROOT maintains
these parts with low cost, by moving their code to new C++ standard
versions, platforms, and compiler versions as needed. Where sensible,
ROOT \href{https://root.cern/doc/master/classTLorentzVector.html}{marks} these parts as legacy, to clearly communicate that now new
adoption of these features should happen, and what the recommended
alternatives are.

\begin{table}[h!]
\centering
\begin{tabular}{ |p{11cm} p{2.5cm}| }
\hline
Lines of code changed per week, 2020 & 2800 \\
Number of contributors, 2020 & Approx. 100 \\
Number of architectures and package managers per ROOT version & 28 \\
New issues (bug reports) in 2020 & 690 \\
Closed issues (bug reports) in 2020 & 660 \\
\hline
\end{tabular}
\caption{Key ROOT development metrics.}
\label{table:metrics}
\end{table}

\subsection{Generational Handover}

For ROOT to be successful in perpetuating its flourishing developments
and converting them into sustainable, trusted software building blocks
for HEP, the upcoming generational handover must succeed. There is a
strong risk that with the retirement of a large fraction of very visible
ROOT developers, expertise or at least visibility and community trust
get retired with them. To reduce this imminent risk considerably, a
small number of currently well-known young experts of ROOT should be
retained until they have a chance to apply for long-term contracts,
succeeding the retired core developers.

\subsection{Cooperation \& Innovation}

Cooperation is essential for ROOT's evolution and a crucial source of
contributions. Fostering cooperation, i.e. an open ROOT project that
newcomers feel invited and welcome to contribute to, is paramount for
this. Being open source and with open discussions and project meetings
is a prerequisite, but in and by itself is not sufficient.

We see how innovation serves as an incentive to contribute. It creates
attention, causes alternatives to be tried by the community, which is an
incentive to benchmark, compare, and improve. The better solutions
generally get integrated into ROOT, where it is of high relevance to the
community and where it matches ROOT's evolution and responsibilities.
Even where innovation does not get integrated because it is outside of
ROOT's scope, it generally backfeeds requirements and ideas, and by
doing so triggers innovation. Cooperation can thus exist on many levels,
from formal ones such as ROOT's current and past CERN knowledge transfer
projects, to informal ones where physicists give presentations to the
ROOT team to share ideas or propose solutions (e.g.~\href{https://indico.cern.ch/event/963454/contributions/4052766/}{Bamboo},
\href{https://indico.cern.ch/event/1042946/}{TTreeIterator},
\href{https://indico.cern.ch/event/1045394/}{CROWN}).

ROOT is working towards making cooperation easier, also on a more formal
level. While this is ongoing, a major milestone in this regard is
expected to be reached in 2023.

\subsection{Creating Opportunities for Externally
Funded Projects}

With ROOT's responsibility towards the community and the ongoing need to
compete with alternative solutions, core development is generally
shouldered by long-term contributors. Many of the peripheral
developments that are still crucial but less time critical can be owned
by specific R\&D projects.

ROOT tries to increase the availability and visibility of sizable, open
R\&D projects that are not on ROOT's critical path. The HEP community
interested in software does not have established mechanisms to
facilitate this; ROOT's past efforts to engage with this community have
had very little successes. ROOT does not yet see the \href{https://hepsoftwarefoundation.org/}{HEP Software
Foundation}
or CERN's \href{https://sidis.web.cern.ch/welcome}{SIDIS} effort
serving as a forum for laboratories or university software R\&D groups
to engage. While investment in software exists, the HEP community might
benefit more from synergies and a more coordinated investment.

In its yearly published \href{https://root.cern/for_developers/program_of_work/}{plan of work},
ROOT will continue to include tasks that are not foreseen to be covered
by existing contributions. The hope is to trigger discussions on these
topics, to identify interested parties, and to engage in more
coordinated R\&D. Having a forum or a set of entities to address this to
would certainly help.

\subsection{Summary of Risks, Functionality Gaps,
Dependencies}

\begin{figure}[H]
\includegraphics[page=1, width=\textwidth]{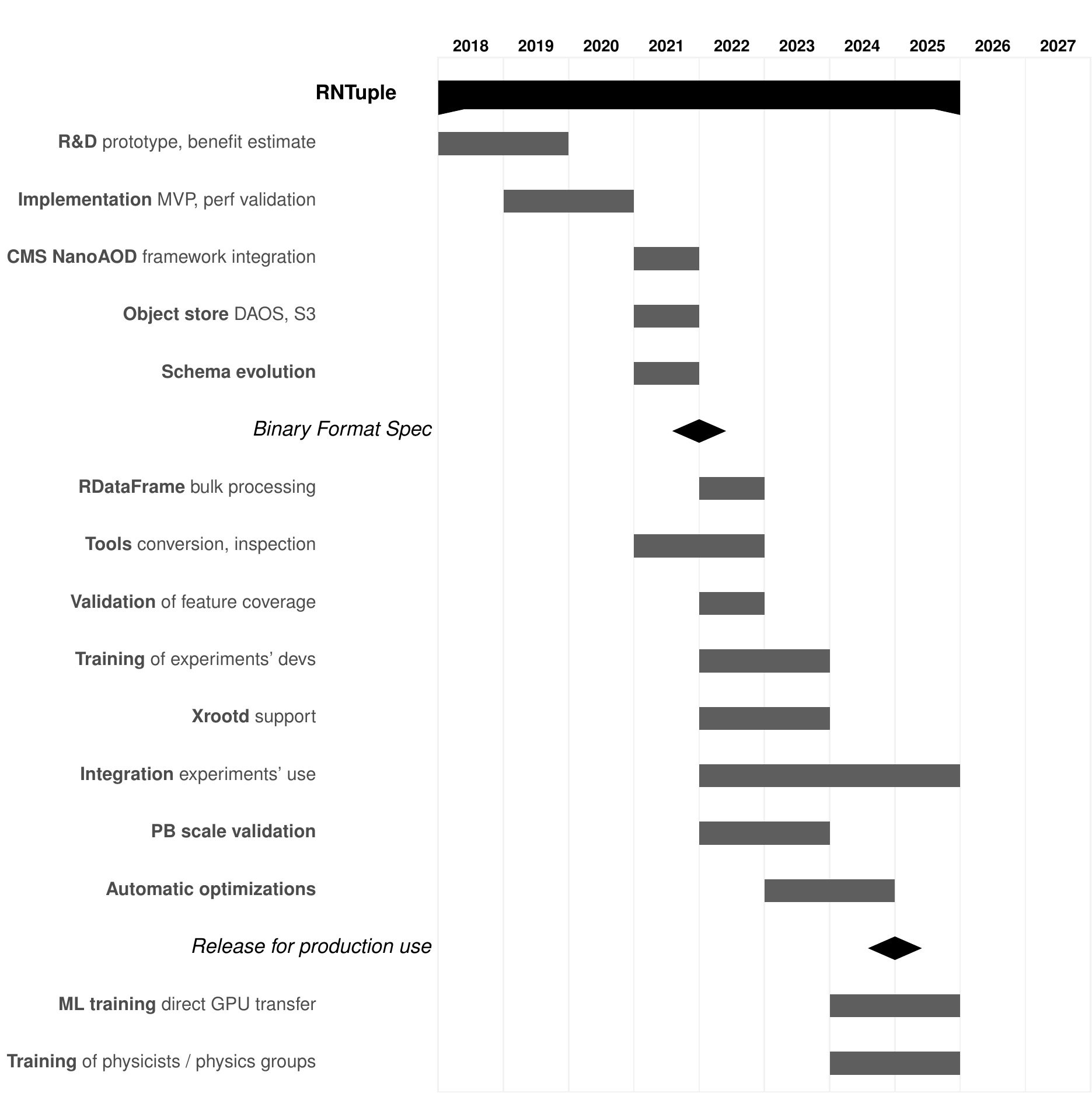}
\caption{Milestones and associated effort: ROOT I/O}
\label{fig:gantt-io}
\end{figure}

Part of ROOT's core responsibilities is the I/O system, as covered in
detail by the \emph{Foundation} part of the ROOT input.
Its milestones and associated effort are shown in Fig.~\ref{fig:gantt-io}.
Other areas and associated risks are as follows.

\subsubsection*{Analysis interfaces: efficient, robust and obvious}

\begin{figure}[H]
\includegraphics[page=2, width=\textwidth]{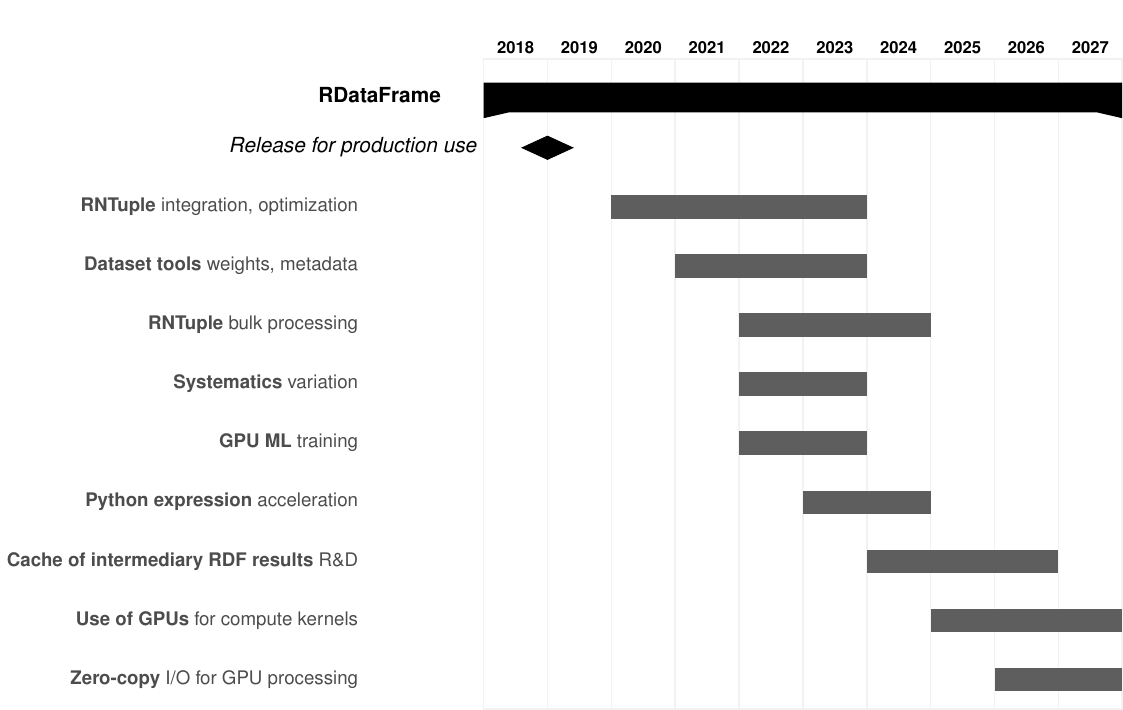}
\caption{Milestones and associated effort: RDataFrame}
\label{fig:gantt-rdf}
\end{figure}

\begin{figure}[H]
\includegraphics[page=4, width=\textwidth]{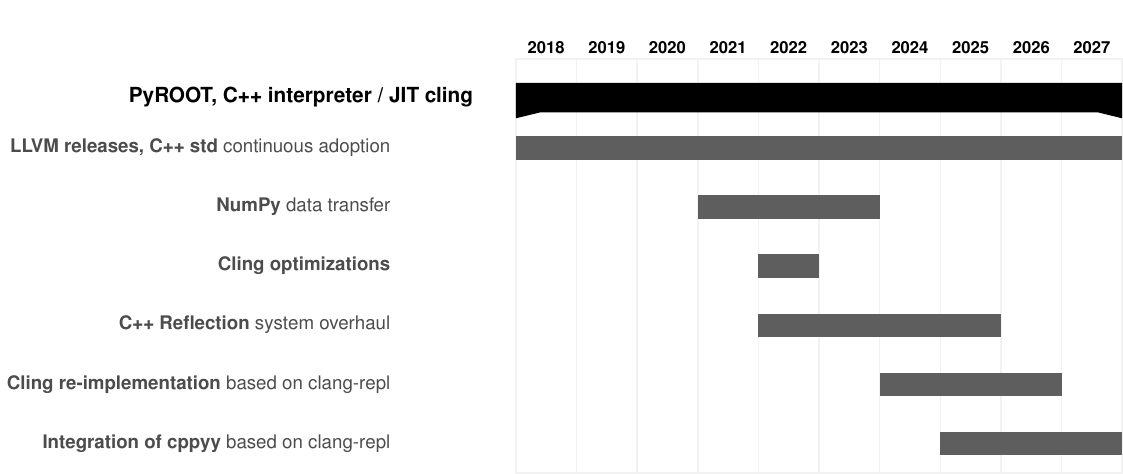}
\caption{Milestones and associated effort: PyROOT, cling C++ Interpreter / JIT}
\label{fig:gantt-interp}
\end{figure}

\begin{figure}[H]
\includegraphics[page=3, width=\textwidth]{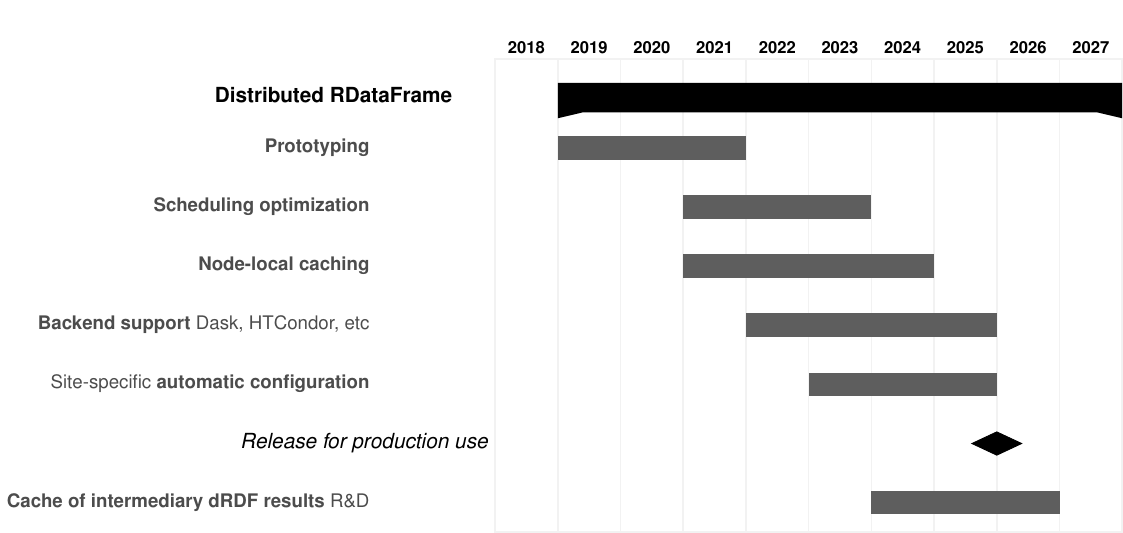}
\caption{Milestones and associated effort: Distributed RDataFrame}
\label{fig:gantt-distrdf}
\end{figure}

\begin{figure}[H]
\includegraphics[page=7, width=\textwidth]{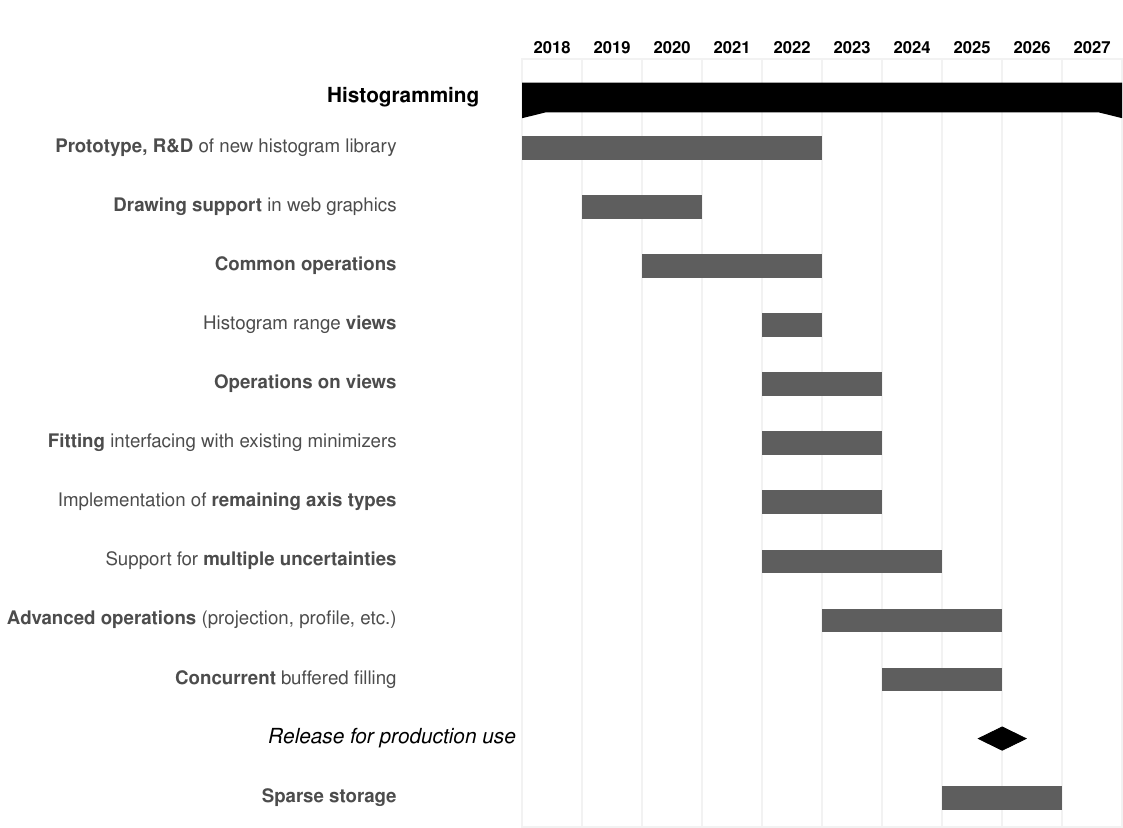}
\caption{Milestones and associated effort: Histogramming}
\label{fig:gantt-hist}
\end{figure}

If evolved properly, ROOT's prime analysis interface, RDataFrame (see Fig.~\ref{fig:gantt-rdf}), can
make use of the impressive uptake from physicists, by guiding them
towards automatically efficient analyses that use all available
resources (network, CPU, GPU). This will enable physicists to work
within a common ecosystem, together build features and tools on top of
RDataFrame, and thus increase and share the benefits of an analysis
environment tailored for HEP. It is paramount to delimit the R\&D
surface: where the open source data analysis community provides
efficient tools that are easy to use, ROOT must invest in
interoperability, with a focus on smooth, efficient usage and highest
possible data transfer bandwidth.

ROOT must invest in interoperability of the expected HL-LHC data format
RNTuple with tools from the open source data analysis ecosystem,
specifically for machine learning (training), NumPy and future (likely
Python-based) data exchange interfaces.

RDataFrame needs to be extended to handle systematic variations within
the same event loop, providing a significant speed-up for the average
analysis. Bulk processing of data must be enabled for RDataFrame, to
benefit from RNTuple's new data layout and to benefit from significantly
higher throughput on CPUs and GPUs.

PyROOT will play an even increasing role in HEP's data analysis
environment. It critically depends on ROOT's C++ interpreter / JIT
compiler cling, and ROOT's type description system, see Fig.~\ref{fig:gantt-interp}.
Future C++
standards, performance and feature bottlenecks must be addressed in
ROOT's type description system. Python-specific adapters to widely used
ROOT interfaces must be implemented to ease their usage also from
Python.

With RDataFrame, ROOT has mostly addressed the issue of "how to write an
analysis accelerated by multithreading". This needs to be extended to a
multi-node environment, currently developed as Distributed RDataFrame, see Fig.~\ref{fig:gantt-distrdf}.
This will reduce the need for the community to develop adapters for
running analyses on clusters such as Spark, for instance by reading ROOT
files in Java. It will make university clusters accessible to
interactive analysis, significantly reducing the turn-around time for
analyses.

ROOT's new histogram library will address usability and performance issues, while providing the feature set expected by HEP analyses, see Fig.~\ref{fig:gantt-hist}.

\subsubsection*{Machine learning models and likelihood functions with ROOT data}

\begin{figure}[H]
\includegraphics[page=5, width=\textwidth]{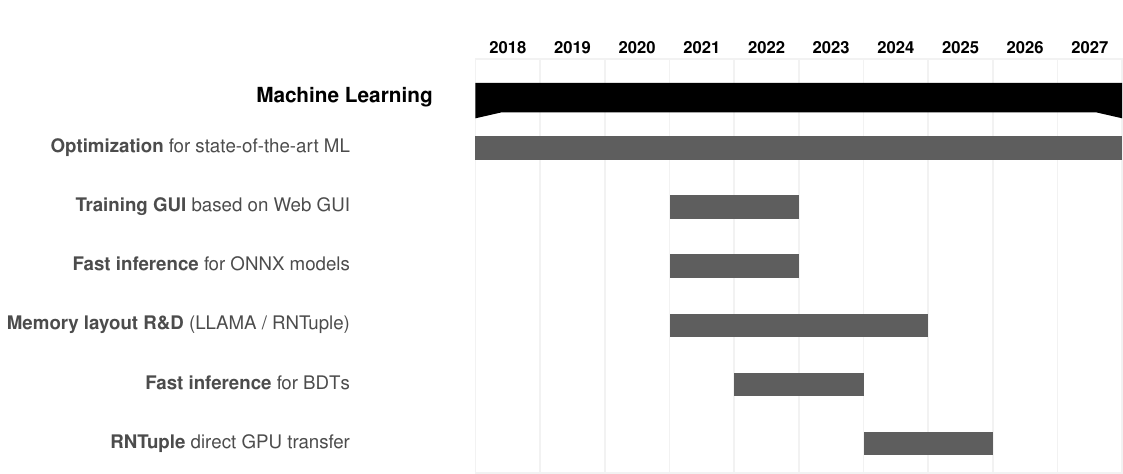}
\caption{Milestones and associated effort: Machine Learning}
\label{fig:gantt-ml}
\end{figure}

\begin{figure}[H]
\includegraphics[page=6, width=\textwidth]{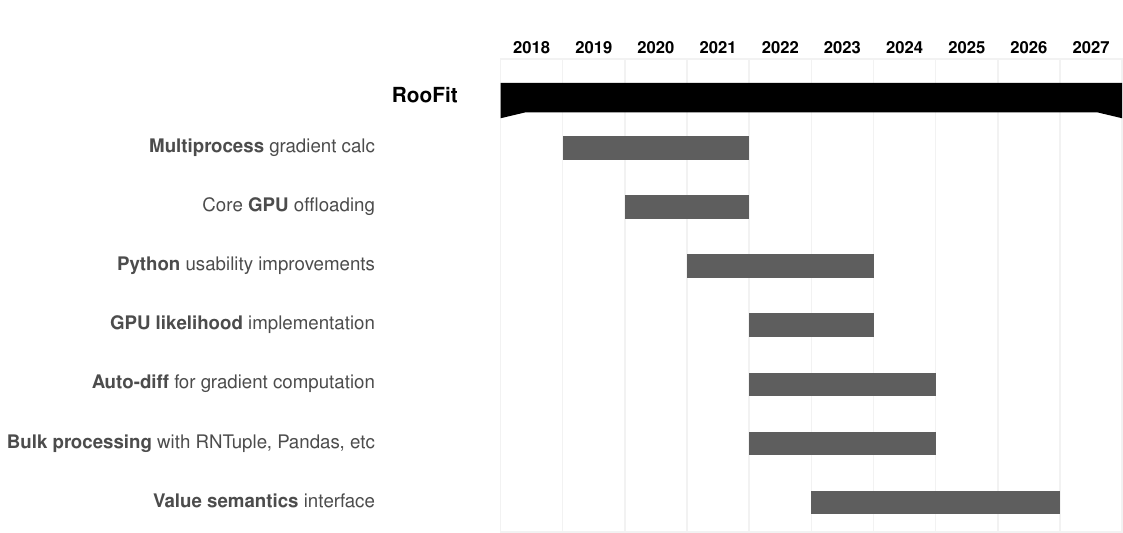}
\caption{Milestones and associated effort: RooFit}
\label{fig:gantt-roofit}
\end{figure}

Many analyses can benefit significantly from direct RNTuple to GPU data
transfer, for instance for machine learning, see Fig.~\ref{fig:gantt-ml}.
This requires work on data
layout and GPU-compatible compression algorithms. Inference from machine
learning models must be simple to use from RDataFrame; results from
RDataFrame must be easily and efficiently usable as ML training input.

RooFit (see Fig.~\ref{fig:gantt-roofit}) needs to continue its renovation for increased efficiency, for
instance by processing arrays of input data also on GPUs. A significant
hurdle of RooFit is the pointer-based interface with implicit ownership
rules; a redesign based on value semantics and thus similar for Python
and C++ is needed for future evolution of RooFit.

\subsubsection*{State-of-the-art visualization}

\begin{figure}[H]
\includegraphics[page=8, width=\textwidth]{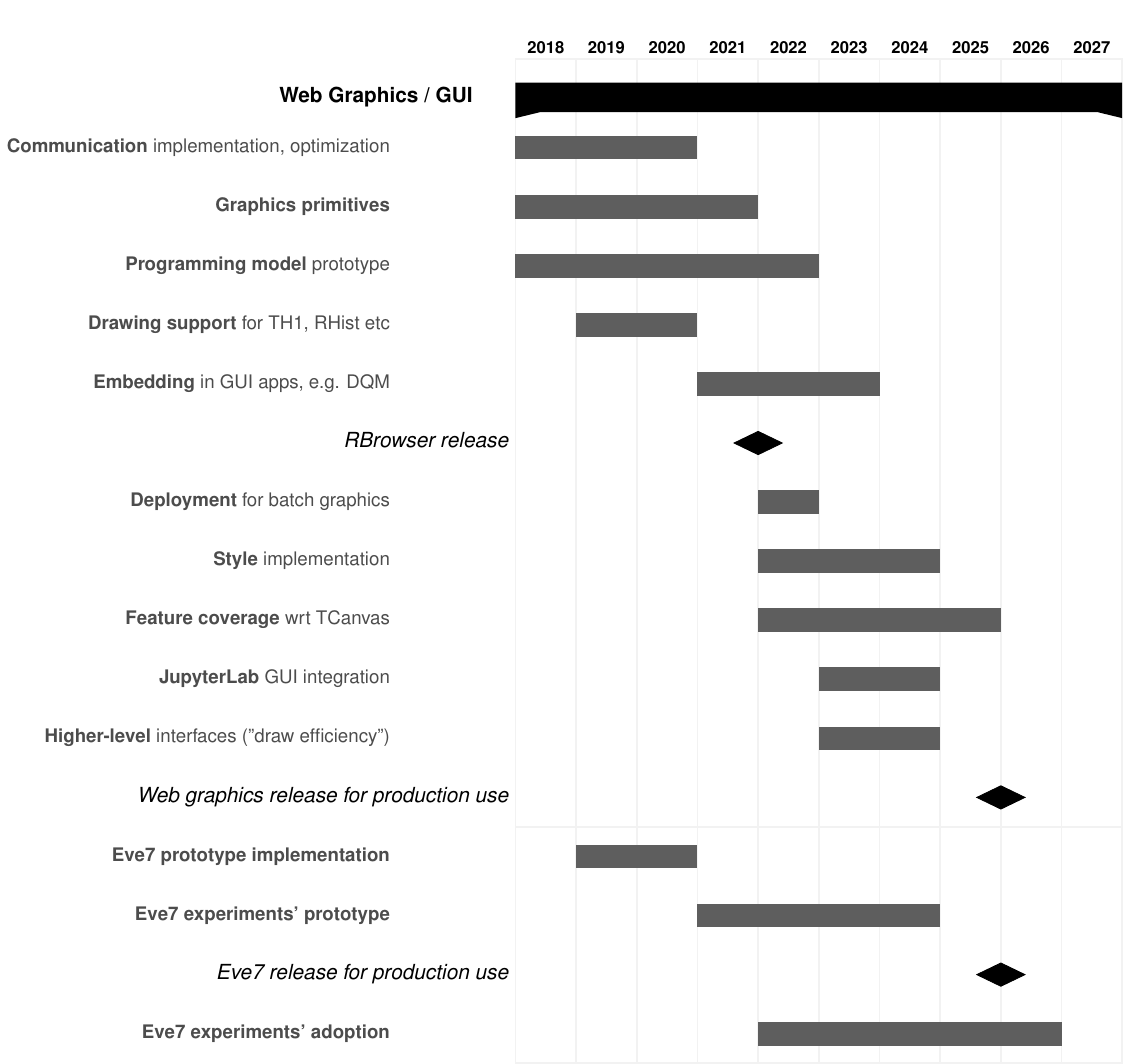}
\caption{Milestones and associated effort: UI and Visualization}
\label{fig:gantt-vis}
\end{figure}

Further investment (see Fig.~\ref{fig:gantt-vis}) will ensure smooth transition from the legacy
graphics and GUI interfaces to the new architecture independent,
web-based graphics and GUI implementations \cite{WebGUI}. For this to succeed, the new
libraries must provide the minimal feature set needed by analyses,
before the legacy libraries will cease to function on commodity analysis
systems, due to these systems deprecating and penalizing X11, GL, Cocoa,
GDI, etc, a process that is currently ongoing.

ROOT must invest in a productivity layer of its very configurable and
currently complex graphics interfaces. This allows graphics to be well
integrated with the rest of ROOT, still highly configurable, but with
defaults that correspond to physicists' expectations. It will guarantee
that the community keeps and commonly evolves its "visual language", for
instance regarding plots showing multiple uncertainty bands,
efficiencies, or higher-dimensional distributions.

ROOT's developing event display seems to satisfy a real community need,
with its high-performance, highly customizable web graphics interface.
ROOT is not aware of any viable alternative with similar functionality
and usability. ROOT expects that this event display will see wide-spread
adoption, if development continues, and is backed by investment in its
constituents such as ROOT's HTTP data transfer (based on ROOT's I/O) and
ROOT's C++ interpreter / JIT compiler cling, and ROOT's web-based
graphics and GUI system.

\begin{landscape}
\footnotesize
\centering
\begin{longtable}{ |p{2.5cm}|p{4cm}|p{1cm}|p{4cm}|p{8cm}| }
\hline
\textbf{Description} & \textbf{Benefits} & \textbf{FTE\footnote{FTE is average per year from 2022 to feature completion; about 50\% senior / 50\% junior developer}} & \textbf{Milestones} & \textbf{Risks, dependencies} \\*
\hline

RNTuple as LH-LHC data format & \textgreater10\% storage reduction; *5
read throughput; reliable error detection; robust user interfaces;
GPU-oriented data layout & 2.5 & 2022: bulk I/O

2022: conversion tools

2023: Xrootd support

2023: PB scale validation

2026: direct GPU I/O & R: delayed features prevent adoption due to lack
of developer effort

R: inefficiencies and trust erosion due to lack of retention of
expertise

R: missed performance improvements due to lack of expertise (software,
storage, network; ROOT and experiments)

R: inability to follow C++ evolution due to lack of resources evolving
I/O and type system support

D: remote I/O libraries (Xrootd, Davix) \\

\hline

RDataFrame as HL-LHC analysis interface & *10 higher physicists'
productivity with obvious analysis interface; O(10) acceleration by
multi-threaded analysis; 2* speedup from transparent optimizations;
effortless migration from TTree to RNTuple & 2 & 2023: dataset weights

2023: systematics variation

2023: GPU ML training

2024: bulk processing

2026: Intermediary result cache

2027: 0-copy GPU processing & D: significant benefits depend on RNTuple
becoming standard analysis format

D: performance and usability of PyROOT affects adoption

D: cling-CUDA implementation for runtime-generated GPU kernels

R: interoperability of RDataFrame with big data analysis ecosystems
affects adoption

R: limited adoption would require physicists' investment to transition
analyses from TTree to RNTuple

R: lack of investment means slower time-to-result for analyses,
triggering migration away from HEP's efficient analysis ecosystem \\

\hline

Distributed RDataFrame & O(10..100) acceleration of analysis
time-to-results (within seconds, "interactive"); smooth migration from
processing analysis locally or on cluster & 1.5 & 2024: node-local
caching

2025: backend support for key schedulers / submission systems

2025: site-specific auto-config & D: requires PyROOT support for
distributing analysis

D: requires analysis written as RDataFrame

R: use of job HEP's main schedulers / submission systems through Python
continues to be supported

R: adoption depends on availability of RDataFrame's Python features \\

\hline

PyROOT, C++ interpreter / JIT compiler cling & Effortless
interoperability with Python ecosystem;

Support of modern C++ standards; efficient language binding accelerating
Python calls into ROOT by factor 2 & 2 & Continuous: integration of new
LLVM releases

2022: cling optimizations

2023: NumPy data transfer from C++ objects

2025: C++ type description overhaul

2026: reimplementation of cling based on clang-repl

2027: integration of cppyy based on clang-repl & D: cppyy as the layer
between ROOT's type description and PyROOT

D: PyROOT depends on cling

D: cling depends on LLVM

R: friction (usability or performance) of using ROOT from Python can
reduce interoperability and usage of ROOT

R: ROOT's type description system cannot represent current C++ standards

R: lack of support of ROOT's type system for cppyy can cause separation
of cppyy from ROOT \\

\hline

Histogramming & Avoids common sources of errors; higher productivity
through simpler, more robust, more efficient interfaces; better
interoperability with modern C++ and Python code & 1.5 & 2022: common
operations

2022: bin ranges

2023: fitting

2023: completion of axis types

2024: multiple uncertainties

2025: concurrent filing

2026: sparse storage & R: adoption requires near feature complete
implementation; little incremental "roll-out" to production use; can
cause design decisions not acceptable by community

R: lack of I/O support for modern C++ features causes extra complexity
in new histograms library

R: lack of interoperable histogram library can cause dispersion over
other libraries / reimplementations \\

\hline

Minimization, Modelling & *10 faster likelihood evaluation;
interoperability & 1.5 & 2023: Python interoperability improvements

2023: GPU likelihood evaluation

2024: gradient computation with auto-differentiation

2026: value semantics & D: PyROOT provides Python interoperability and
efficient data transfer

D: CUDA / GPU programming model

D: ciad for automatic differentiation

R: Limited physics reach of HL-LHC data due to performance-induced
limitation of model complexity \\

\hline

Machine Learning & *10 faster ML training data throughput from ROOT
files & 2 & Continuous: support for using state-of-the-art ML models
with ROOT data

2022: fast inference of ONNX models

2023: fast inference for BDTs

2025: direct GPU transfer for ML training of RNTuple data

& D: ML ecosystem

D: CUDA / GPU programming model

R: lack of efficient interoperability causes data conversion, additional
storage use, and potentially reduction in physics reach (memory-limited
input data for model training) \\

\hline

Web-based Graphics, GUI & Platform-independent visualization code with
reduced need for platform UI experts; integration in GUI applications;
remote-graphics capabilities; physicists produce graphics more
effectively & 1.5 & 2023: embedding in GUI applications

2024: style definition

2024: higher-level graphics interface

2025: feature equality with TCanvas & D: OpenUI5 for Web-GUI

D: use of D3.js, three.js for JavaScript visualization in 2D, 3D

D: Chrome Embedded Framework (directly or through Qt); browser
availability and configurability

D: civetweb

R: security implications of web-based graphics can jeopardize
adoption \\

\hline

Event Display & Improved analysis through interactive inspection of
event selection; improved outreach / communication; accessible detector
design / simulation & 1 & 2024: CMS (and likely others such as Mu2e)
event display prototype & D: Web-based Graphics, GUI

\\

\hline
\caption{Summary of benefits, priorities, efforts, and risks of main work areas.}
\label{table:workareas}
\end{longtable}
\end{landscape}

For ROOT, RNTuple has the highest priority and is most time critical, as
adoption depends on production readiness well before the start of
HL-LHC. Analysis interfaces (RDataFrame, RooFit, machine learning and
Python interoperability, histogramming) are the second most important
development, defining the future of ROOT-based HEP analysis, with a
significant benefit for the efficiency of physicists and the impact of
delivered luminosity. Progress is defined by a combination of these
priorities and available expertise and developer effort. In general,
ROOT must provide continuous effort for support and maintenance of all
these fields, which in turn guarantees a sustained development progress
in all these fields.

\section{Summary}

Several years ago, ROOT had identified the HL-LHC as a welcome timeline
to reinvent itself. Today's ROOT can still function as it did ten or
twenty years ago, thanks to its backward compatibility. But many physics
analyses have moved away from that, embracing ROOT's new features that
come with a much increased usability and efficiency. This endeavor has
attracted many R\&D contributions, helping with fundamental innovation
and fast progress in ROOT's evolution. All of ROOT's core areas have
benefited from this investment: the I/O system, likelihood evaluation,
machine learning, data analysis interfaces including distributed
analysis, Python bindings and ROOT's "Python personality", ease of
installation and deployment.

Part of ROOT's evolution speed is due to technology advances being
accessible to other parts of ROOT: the web graphics system and the event
display benefit from ROOT's JavaScript interface JSROOT, the I/O
subsystem and the interpreter; RooFit's vectorization efforts benefit
from experience in vectorizing ROOT's fitting algorithms; the
development teams shares commonly accrued expertise on writing
high-performance, highly concurrent code; ROOT's Python-specific
interfaces benefit from in-house experience of data transfer into Python
other features of PyROOT and the C++ interpreter / JIT compiler. ROOT is
itself an ecosystem, re-using its own innovations to multiply their
effect.

At the same time, ROOT is embedded in an ecosystem of tools built on top
of ROOT, or bridging into other data science ecosystems. ROOT works on
providing these bridges itself, arguing that efficiency and separation
of concerns are best taken care of by experts who can guide the analysis
physicists in their use of for instance ROOT data with machine learning
tools. ROOT's technologies and the continuing relevance of C++ will
likely allow ROOT to also satisfy interoperability needs for the next 20
years.

Yet, ROOT cannot reinvent itself without providing sustainability for
those new features, which in turn requires long-term commitment of
developer resources, and the ability of these developers to also invest
significantly in work not related to R\&D. The currently ongoing peak of
innovation will need to transition into an era of optimizations and
support, during the HL-LHC data taking years. To guarantee this,
retaining expertise and community trust is paramount.

\section{Conclusion}

ROOT is not only a software project: it is a team of technical experts
and innovators, communicators with the HEP community, people that this
community trusts and knows to rely on. At age 28, ROOT is currently
reinventing itself, to benefit from the team's and the community's
experience and support, and to adjust to tomorrow's challenges and
requirements, in time for the HL-LHC.

ROOT's evolution is driven by a rich bouquet of innovation with steady
progress over the last couple of years. The goal is to bring ROOT to the
level of usability, efficiency, robustness, and community trust required
for HL-LHC analyses. Efficiency improvements in orders of magnitude have
been seen across the board; adoption by physicists has been impressively
rapid, likely because of the features' quality, but also due to
increased efforts invested in communication of these new features, and
involvement of the analysis community already during the development of
the new features, following ROOT's tradition.

Even though ROOT is working on filling several feature, performance, and
productivity gaps, it needs to be able to do so in a sustainable way:
today's new features are tomorrow's source of bugs. ROOT relies on
continued assistance from the community and contributors to provide user
support and maintenance for HL-LHC. It especially needs this assistance
to transition from the current active R\&D phase into a support phase,
with retained expertise.

\bibliographystyle{unsrt}
\bibliography{bibliography}

\end{document}